\documentclass[a4paper,twocolumn,english,aps,prb,amsmath,amssymb]{revtex4-1}
\pdfoutput=1

\usepackage[utf8]{inputenc}
\usepackage[english]{babel}
\usepackage{amsmath,amsfonts,amsthm,amssymb}
\usepackage{mathtools}
\usepackage{graphicx}
\usepackage{color}
\usepackage{bm}
\usepackage[inline]{enumitem}
\usepackage{array}

\newcommand{\ep}{\epsilon}
\DeclareMathOperator{\sgn}{sgn}

\begin{document}

\title{Modeling the Excitation of Graphene Plasmons in Periodic Grids of Graphene Ribbons: An Analytical Approach}

\author{P. A. D. Gon\c{c}alves$^{1,2,3}$}
\email{Electronic address: padgo@fotonik.dtu.dk}
\author{E. J. C. Dias$^{1}$}
\author{Yu. V. Bludov$^{1}$}
\author{N. M. R. Peres$^{1}$}
\email{Electronic address: peres@fisica.uminho.pt}

\affiliation{$^{1}$Department of Physics and Center of Physics, University of Minho, PT-4710-057, Braga, Portugal}

\affiliation{$^{2}$ Department of Photonics Engineering, Technical University of Denmark, DK-2800 Kgs. Lyngby, Denmark}

\affiliation{$^{3}$Center for Nanostructured Graphene (CNG), Technical University of Denmark, DK-2800 Kgs. Lyngby, Denmark}

\date{\today}

% -------------------------------------------------------------
% #                          Abstract                         #
% -------------------------------------------------------------

\begin{abstract}
We study electromagnetic scattering and subsequent plasmonic excitations in periodic grids of graphene ribbons. To address this problem, we develop 
an analytical method to describe the plasmon-assisted absorption of electromagnetic radiation by a periodic structure of 
graphene ribbons forming a diffraction grating for THz and mid-IR light. The major advantage of this method lies in its ability to 
accurately describe the excitation of graphene surface plasmons (GSPs) in one-dimensional (1D) graphene gratings without the use of both  
time-consuming, and computationally-demanding full-wave numerical simulations. We thus provide analytical expressions for the reflectance, 
transmittance and plasmon-enhanced absorbance spectra, which can be readily evaluated in any personal laptop with little-to-none programming. 
We also introduce a semi-analytical method to benchmark our previous results and further compare the theoretical data with spectra taken from experiments, 
to which we observe a very good agreement. 
These theoretical tools may therefore be applied to design new experiments and cutting-edge nanophotonic devices based on graphene plasmonics.
\end{abstract}

\maketitle
%\tableofcontents

% *************************************************************
% ::                        INTRODUCTION                     ::
% *************************************************************
\section{Introduction}

Nowadays, photonics --- dubbed ``the science of light'' --- is one of the branches of the physical sciences 
with most impact in our daily lives. It is concerned with the study and manipulation 
of light (photons) in a manifold of fundamental and technological landscapes. 
Recently, the ``nano-revolution'' under way has led to the miniaturization of electronics. However, in what regards 
electromagnetic (EM) radiation, such miniaturization is limited by the length-scale defined by the wavelength of the employed light (known 
as the diffraction-limit). In this context, 
plasmonics\cite{MaradudinModern2} has been regarded as the most promising candidate to bring EM fields 
to the nanoscale.\cite{Barnes,Gramotnev:2010,plasmExtreme,difractionfree,Ozbay} 
Plasmonics is a branch of photonics which deals with quasiparticles known as plasmon-polaritons.\cite{MaradudinModern2,Toropov} 
Surface plasmon-polaritons (SPPs) are electromagnetic surface waves coupled to collective excitations of the free electrons in conductors. When 
these hybrid excitations occur in 
conducting nanostructures --- 
such as nanoparticles\cite{Pelton,NPs} or engineered metamaterials\cite{Billings,Kivshar} ---, the corresponding non-propagating 
plasmon-polaritons are generally coined as localized 
surface plasmons (LSPs).\cite{MaradudinModern2} Perhaps the most alluring property of plasmons is that they exhibit large field-enhancements 
and deep subwavelength confinement of EM fields, thereby circumventing the 
diffraction limit of conventional optics.\cite{Barnes,Gramotnev:2010,plasmExtreme,difractionfree} For 
this reason, plasmonics has been considered the ultimate pathway to manipulate light-matter interactions at the nanometer scale.

Very recently, graphene\cite{Geim09,RMP81,C4NR01600A} --- a two-dimensional (2D) crystal made up of carbon atoms arranged in a honeycomb lattice --- has emerged as 
a promising plasmonic material, benefiting from this material's remarkable electronic 
and optical properties.\cite{RMP81,C4NR01600A,nl102824h,nphto_GPhot} 
Doped graphene is capable of supporting 
SPPs --- graphene surface plasmon-polaritons (GSPs)\cite{nphReview,AbajoACSP,Xiao2016,ACSgp,nlgp,Primer,C4NR03143A,GoncalvesPeres} --- in the THz and mid-IR 
spectral range. These possess tantalizing properties, outperforming traditional noble-metal plasmonics, in that spectral window, in terms of mode confinement, and are predicted to 
suffer from relatively low losses when compared to customary three-dimensional (3D) metals.\cite{nlgp,AbajoACSP,Xiao2016,MRS:8669489} In addition, 
graphene plasmons have yet another 
key advantage: the ability of being actively tunable by means of electrical gating or chemical doping. This feature 
constitutes a major improvement over conventional metal-based plasmonics\cite{MaradudinModern2,Toropov} (where tunability is usually limited by 
the geometry and composition of the system, and therefore it is fixed), and constitutes a sought-after characteristic for active nanophotonic 
devices and/or circuitry based on graphene plasmons.
Indeed, a plethora of proof-of-concept, application-oriented experiments have already demonstrated the capabilities of GSPs to 
deliver extremely sensitive biochemical sensors,\cite{Rodrigo10072015,ACSphot6b00143,am25,goldplasmonsgraphene,PCCP15} 
surface-enhanced Raman scattering (SERS),\cite{XLing,PNAS109,Zhao,CYLiu,nl13,apl101Raman} 
polarizers,\cite{TEpol,NJPChristian} optical modulators\cite{Sun16,apl101,Renwen,Sensale-Rodriguez:13} and 
photodetectors.\cite{nl13Photo,ncom2,C5NR08677A,acs.nl.5b02051,koppensphoto}

Such achievements are particularly notable in the light that optical excitation of graphene plasmons was only achieved as recently 
as in 2011 by Ju \textit{et al.}, using periodic arrays of graphene microribbons.\cite{NatNano} That foundational publication paved the way 
for the emergence of many experimental and theoretical works that soon followed, thereby establishing the field of 
graphene plasmonics.\cite{GoncalvesPeres,nphReview,AbajoACSP,Xiao2016,nlgp,ACSgp,Primer,C4NR03143A} As of today, GSPs have been realized 
in a number of systems, ranging from patterned grids of 
graphene ribbons,\cite{NatNano,nphoton7,APL105,Rodrigo10072015,ACSphot6b00143,NJPChristian,apl101,nnano7,luxmoore14,C5NR05175D}  
disks,\cite{NJP14,nnano7,ACS7,ZFnl14,nl14XZ} and rings, \cite{ACS7,NJP14} 
periodic anti-dot lattices,\cite{Optica2,nl14XZ,PlasmonicCrystals} resonators,\cite{nl13Brar,nlAtwater}  
hybrid graphene/metal nano-antennas,\cite{nl13Yao,nl13Photo,ACS6Halas,ncom2,Science344}  
among others.\cite{nl13Gao,APL102,nl11,nl12mag2,nonlinearplasmons,nl_3waveMix,nanoImBasov,nanoImICFO,Woessner2015,C5NR01056J}

In the heart of plasmonics lies the fact that freely-propagating EM radiation cannot 
couple directly to plasmons owing to the momentum mismatch between plasmons and photons of the same frequency. However, the property 
that the plasmon's wavevector is larger than the wavevector of light of the same frequency is exactly what enables extreme localization of light 
into subwavelength volumes. For extended graphene, these volumes can be about $\alpha^3 \approx 10^{-6}$ times smaller (where $\alpha$ denotes 
the fine-structure constant) than the volume characterized by 
the free-space light's wavelength (i.e. $\lambda_0^{-3}$). Typical strategies to couple light to graphene plasmons involve the 
patterning of pristine graphene into 
gratings and related nanostructures,\cite{NatNano,nphoton7,APL105,Rodrigo10072015,NJPChristian,apl101,nnano7,luxmoore14,C5NR05175D,ACSphot6b00143,NJP14,nnano7,ACS7,ZFnl14,nl14XZ,Optica2,nl14XZ,PlasmonicCrystals}  
the use of dielectric gratings,\cite{nl13Gao,APL102} light scattering from a conductive tip,\cite{nanoImBasov,nanoImICFO,Woessner2015,C5NR01056J} 
and even non-linear three-wave mixing.\cite{nonlinearplasmons,nl_3waveMix}

In this context, the utilization of periodic grids of graphene 
ribbons % (and disks for that matter) 
--- fabricated by patterning an otherwise continuous graphene sheet --- 
has been one of the most popular setups to realize graphene plasmons with energies 
from the THz up to the mid-IR regime, which can be tailored either by varying the size of the ribbons or 
by tuning the concentration of charge-carriers in graphene (and thus the Fermi level). 
Under this scheme, the array of graphene ribbons effectively acts as a diffraction grating for EM radiation impinging 
on the system (e.g. from a laser), producing scattered waves which carry momenta in multiples of the reciprocal 
lattice vector, $G=2\pi/L$ (where $L$ is the grating 
period), thus overcoming the above-mentioned kinematic constraint.
The reason the use of ribbon arrays to couple light to GSPs has been so predominant is essentially 
two-fold (apart from being easily attainable with current fabrication technologies): it enable us to overcome the 
momentum mismatch between light and GSPs; and it renders a stronger (composite) plasmonic response than one would get from a 
single graphene ribbon (also, in this latter case, instead of well-defined diffracted orders, the scattered waves 
would transport a continuum of momenta).

In this work we develop an analytical framework describing the interaction of EM radiation 
with periodic grids of micro- and nano-sized graphene ribbons. The main motivation driving this work was 
to deliver a simple and transparent theoretical tool capable of explaining the plasmon-induced spectra measured 
in experiments that did not involve the use of computationally-heavy and time-consuming numerical 
simulations. Here, we provide simple closed-form expressions for the reflectance, transmittance and 
absorbance spectra of THz and mid-IR light through graphene patterned into 
ribbons. These spectra may then be used to design or model experiments with graphene plasmons in the laboratory, 
by simply evaluating an analytical expression. The coupling between graphene plasmons and surface optical (SO) 
phonon modes of a SiO$_2$ substrate is also considered, and we observe a reconstruction of the 
polaritonic spectrum owing to the hybridization of GSPs with SO phonons of the underlying polar substrate.
We further introduce a semi-analytical technique developed elsewhere\cite{JPCM24,PolCrysPRB85,JOpt15,Primer,GoncalvesPeres} 
to benchmark our analytical theory. Finally, we compare the outcomes of both frameworks against actual experimental data 
and demonstrate their ability to describe plasmonic excitations in periodic gratings of graphene ribbons.

% ================================================== // ==================================================

\section{Theory}

% *************************************************************
% ::         THEORY AND RESULTS: ANALYTICAL MODEL            ::
% *************************************************************

\subsection{Analytical Method}

\begin{figure}[h!]
\centering
 \includegraphics[width=\the\columnwidth]{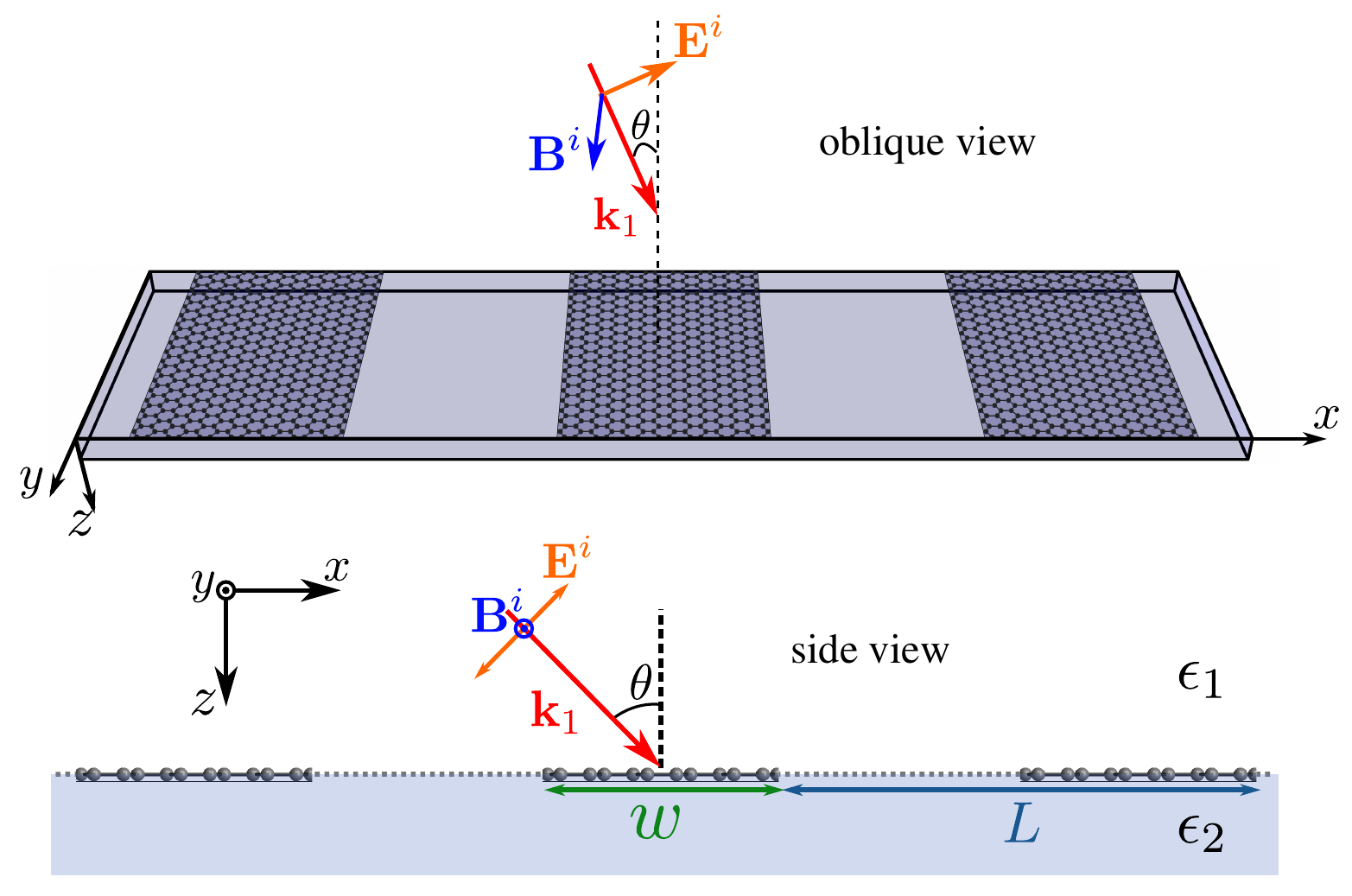}
  \caption[Periodic grid of graphene ribbons]{
  Monochromatic $p-$polarized plane-wave impinging on a grid of graphene ribbons (not to scale) arranged in a grating-like configuration. 
  The ribbons are sitting in the plane defined by $z=0$. The structure is periodic, with period $L$, and the width of the 
  graphene ribbons is defined by $w$. The system is encapsulated between a top insulator with 
  relative permittivity $\epsilon_1$ (for $z<0$), and a dielectric substrate with relative 
  permittivity $\epsilon_2$ (for $z>0$).}\label{fig:graphene_grid_FV}
\end{figure}

We consider the scattering of EM radiation by a 1D periodic grid of 
graphene ribbons of width $w$. For the sake of simplicity, the ribbons are assumed to possess infinite length in the longitudinal direction. In such an arrangement, 
the graphene grid behaves like a diffraction grating for EM waves. The period of the grating is denoted by $L$ hereafter, and the system 
is assumed to lie in the plane defined by $z=0$, being cladded between two dielectric media 
with relative permittivities $\epsilon_1$ (for $z<0$) and $\epsilon_2$ (for $z>0$) --- see 
Fig. \ref{fig:graphene_grid_FV}. In what follows, we assume ribbons whose widths are $\sim 100$ nm or larger, so that 
the actual edge termination of the graphene ribbons and finite-sized effects are not important, 
and therefore a classical electrodynamics framework suffices \cite{ACS6,PRB90:TC}.

We consider a $p$-polarized monochromatic plane-wave impinging on the grid of graphene ribbons at an angle $\theta$. For such polarization, 
the incident EM fields read
\begin{align}
 \mathbf{B}^i (\mathbf{r},t) &= B^i_0\ e^{i( \mathbf{k}_1 \cdot \mathbf{r} - \omega t)}\ \mathbf{\hat{y}}\ , \label{eq:B_inc}\\
 \mathbf{E}^i (\mathbf{r},t) &= \left( E^i_{0,x}\ \mathbf{\hat{x}} + E^i_{0,z}\ \mathbf{\hat{z}} \right) 
 e^{i( \mathbf{k}_1 \cdot \mathbf{r} - \omega t)}\ , \label{eq:E_inc}
\end{align}
where the wavevector of the incoming wave is defined as 
$\mathbf{k}_1 = k_x\ \mathbf{\hat{x}} + k_z\ \mathbf{\hat{z}}$, with $k_x = \sqrt{\ep_1} k_0 \sin\theta$ and 
$k_z = \sqrt{\ep_1} k_0 \cos\theta$, where $k_0 = \omega/c$. 
Naturally, the field amplitudes $B^i_0$, $E^i_{0,x}$ and $E^i_{0,z}$ are connected via Maxwell's equations, 
which establish the relations $E^i_{0,x} = \frac{c^2 k_z}{\omega \ep_1} B^i_0$ and $E^i_{0,z} = -\frac{c^2 k_x}{\omega \ep_1} B^i_0$.
Furthermore, due the periodicity of the grid, one may write the reflected magnetic field in the form of a Bloch-sum 
(also termed as Fourier-Floquet decomposition),
\begin{equation}
 \mathbf{B}^r (\mathbf{r}) = \sum_{n=-\infty}^{\infty} r_n\ e^{i( q_n x - \kappa^{-}_{z,n} z )}\ \mathbf{\hat{y}}\ , \label{eq:B_ref}
\end{equation}
where an implicit time-dependence of the usual form $e^{-i\omega t}$ is assumed henceforth, and where the wavevectors of the Bloch modes are defined 
as
\begin{equation}
 q_{n} = k_x + nG = k_x + n 2\pi/L\ , \label{eq:qx_ref}
\end{equation}
where $G=2\pi/L$ is the primitive vector of the reciprocal lattice. In addition, note that
$
 \ep_1 k_0^2 = q^2_n + \left( \kappa^{-}_{z,n} \right)^2
$
as determined from Maxwell's equations. Likewise, the field transmitted across the graphene grating may also be casted as a Bloch-sum, reading
\begin{equation}
 \mathbf{B}^t (\mathbf{r}) = \sum_{n=-\infty}^{\infty} t_n\ e^{i( q_n x + \kappa^{+}_{z,n} z )}\ \mathbf{\hat{y}}\ , \label{eq:B_trans}
\end{equation}
where $\left( \kappa^{+}_{z,n} \right)^2 = \ep_2 k_0^2 - q^2_n$. 
As in the case of the incident fields, we can make use of Maxwell's curl equation $\nabla \times \mathbf{B} = -i\omega \ep/c^2 \mathbf{E}$ to 
write out the corresponding reflected and transmitted electric fields from Eqs. (\ref{eq:B_ref}) and (\ref{eq:B_trans}). This procedure leads to
\begin{align}
 \mathbf{E}^r (\mathbf{r}) &= -\frac{c^2}{\omega \ep_1} \sum_{n=-\infty}^{\infty} r_n
 \left[ \kappa^{-}_{z,n} \mathbf{\hat{x}} + q_n \mathbf{\hat{z}} \right] 
 e^{i( q_{n} x - \kappa^{-}_{z,n} z )}\ , \label{eq:E_ref} 
\\
 \mathbf{E}^t (\mathbf{r}) &= \frac{c^2}{\omega \ep_2} \sum_{n=-\infty}^{\infty} t_n
 \left[ \kappa^{+}_{z,n} \mathbf{\hat{x}} - q_n \mathbf{\hat{z}} \right] 
 e^{i( q_{n} x + \kappa^{+}_{z,n} z)}\ , \label{eq:E_trans}
\end{align}
respectively. At this stage, the coefficients $r_n$ and $t_n$ are still unknown. In order to determine them, one must impose the appropriate 
boundary conditions of the problem. To that end, we employ the first boundary condition stating that the $x$-component of the electric field 
above and below the graphene grid must be continuous, $\mathbf{\hat{x}} \cdot \left( \mathbf{E}^i + \mathbf{E}^r - \mathbf{E}^t \right)|_{z=0} = 0$, 
that is
\begin{equation}
 k_z B_0^i e^{i k_x x} - \sum_{n=-\infty}^{\infty} r_n \kappa^{-}_{z,n} e^{i q_{n} x} = 
 \frac{\ep_1}{\ep_2} \sum_{l=-\infty}^{\infty} t_l \kappa^{+}_{z,l} e^{i q_{l} x}\ .
\end{equation}
Multiplying the previous expression with a basis function, $e^{-i q_m x}$, and integrating over the unit cell, yields
\begin{equation}
 r_m = \frac{k_z}{\kappa^{-}_{z,m}} B_0^i \delta_{m,0} - \frac{\ep_1}{\ep_2} \frac{\kappa^{+}_{z,m}}{\kappa^{-}_{z,m}} t_m
 \ , \label{BC1:rm_Bi0_tm}
\end{equation}
which links the Bloch coefficients $r_m$ and $t_m$ (and $B_0^i$ for that matter).
Moreover, according to Ohm's law, the electric fields produce a current given by
$
 \mathbf{J} = \sigma(x) \mathbf{\hat{x}} \cdot \mathbf{E}^t |_{z=0} \mathbf{\hat{x}}
$ 
, which reads
\begin{equation}
 J_x(x) = \frac{\sigma(x) c^2}{\omega \ep_2}
 \sum_{n=-\infty}^{\infty} \kappa^{+}_{z,n} t_n e^{i q_{n} x}\ , \label{eq:current_Ohm}
\end{equation}
where $\sigma(x)$ is the position-dependent conductivity of graphene, which in the unit cell can be written as 
$\sigma(x) = \sigma(\omega) \Theta(w/2 - |x|)$. In this expression, $\sigma(\omega)$ is the dynamical conductivity of a 
graphene ribbon (here assumed to be bulk-like) and $\Theta(x)$ denotes the Heaviside step function.\cite{AS}

We now introduce a central assumption into our analytic method, which is the validity of the \emph{edge condition}.\cite{Barkeshli}  
This condition states that the current perpendicular to a sharp edge --- 
such that of a graphene ribbon --- should be proportional to the square-root of the distance to the edge, $\rho$, that is, 
$J_x(\rho) \propto \sqrt{\rho}$. Our assumption here is that in the regime where $k w <1$ one 
can interpolate the current by an expression that incorporates the edge condition at both edges of each ribbon, e.g. $x=\pm w/2$, simultaneously. 
Therefore, this ansatz allows us to write the current within a ribbon in the unit cell as
\begin{equation}
 J_x (x) = \chi e^{i k_x x} \sqrt{w^2/4 - x^2} \Theta(w/2 - |x|)\ , \label{eq:Jx_ansatz}
\end{equation}
where $\chi$ is a coefficient to be determined.
As a first step towards the determination of the coefficient $\chi$, we now argue that Eqs. (\ref{eq:current_Ohm}) and (\ref{eq:Jx_ansatz}) 
must give rise to the same induced current (since they represent the same physical quantity). Hence, one may write the following relation 
(in the unit cell)
\begin{equation}
 \chi e^{i k_x x} \sqrt{w^2/4 - x^2} \Theta(w/2 - |x|) = \frac{\sigma(x) c^2}{\omega \ep_2}
 \sum_{n=-\infty}^{\infty} \kappa^{+}_{z,n} t_n e^{i q_{n} x}\ , \label{eq:equating_Jx}
\end{equation}
which, after multiplying by a basis function, $e^{-i q_m x}$, and integrating over the unit cell, produces
\begin{equation}
 \chi \frac{L}{4m} J_1(m\pi w /L) = \frac{\sigma(\omega) c^2}{\omega \ep_2} 
 \sum_{n=-\infty}^{\infty} \kappa^{+}_{z,n} t_n \frac{\sin\left( [n-m] \pi w/L \right)}{[n-m] \pi w/L}
\ , \label{rel:chi_tm}
\end{equation}
where $J_1(x)$ is the 1st-order Bessel function of the first kind \cite{AS}. This 
expression defines the coefficient $\chi$ in terms of the Bloch-amplitudes 
$t_m$, and whose combination with Eq. (\ref{BC1:rm_Bi0_tm}) 
connects the coefficients $\chi$, $r_m$ and $t_m$. In order to close the system 
of equations, we require another expression relating these quantities. Such ``extra'' equation is the other boundary condition holding for this system, 
in particular, the discontinuity of the magnetic field across the graphene grid due to the presence of the surface current induced by 
the electric field, i.e. 
$
 \mathbf{\hat{z}} \times \left( \mathbf{B}^t - \mathbf{B}^r - \mathbf{B}^i \right)|_{z=0} = \mu_0 J_x \mathbf{\hat{x}}
$
which, after applying the same operations that led to Eqs. (\ref{BC1:rm_Bi0_tm}) and (\ref{rel:chi_tm}), gives
\begin{equation}
 B_0^i \delta_{m,0} = t_m - r_m + \mu_0 \chi \frac{w}{4m} J_1(m\pi w /L)
\ , \label{BC2:rm_Bi0_tm_chi}
\end{equation}
thereby closing the system.
Finally, the combination of Eqs. (\ref{BC1:rm_Bi0_tm}) and (\ref{BC2:rm_Bi0_tm_chi}) allows us to write the $t_m$'s as
\begin{equation}
 t_m = \frac{\ep_2 \kappa^{-}_{z,m}}{\ep_1 \kappa^{+}_{z,m} + \ep_2 \kappa^{-}_{z,m}}
 \left[ 2 B_0^i \delta_{m,0} - \mu_0 \chi \frac{w}{4m} J_1(m\pi w /L) \right]
\ , \label{eq:tm_in_terms_of_chi}
\end{equation}
which, after using Eq. (\ref{rel:chi_tm}) endows us an expression for the coefficient $\chi$ (from which the Bloch amplitudes 
$t_m$ and $r_m$ directly follow), that is
\begin{equation}
 \chi = \frac{2 \kappa^{+}_{z,0} \kappa^{-}_{z,0}}{\ep_1 \kappa^{+}_{z,0} + \ep_2 \kappa^{-}_{z,0}} 
 \frac{\sigma(\omega) c^2}{\omega} \frac{B_0^i}{\Lambda(\omega)}
\ , \label{res:chi_result}
\end{equation}
where the quantity $\Lambda(\omega)$ is defined as
\begin{equation}
 \Lambda(\omega) = \frac{w}{4} \sum_{n=-\infty}^{\infty} \frac{1}{n} J_1(n\pi w /L) 
 \left[ 1 + \frac{\sigma(\omega)}{\omega \ep_0} \frac{ \kappa^{+}_{z,n} \kappa^{-}_{z,n}}{\ep_1 \kappa^{+}_{z,n} + \ep_2 \kappa^{-}_{z,n}} \right]
 \ . \label{def:Lambda}
\end{equation}
It should be stressed that the sum in the previous expression needs to be judiciously performed, 
since it is a sum with alternating signs (check appendix \ref{appx:sum} for details).

Therefore, Eqs. (\ref{BC1:rm_Bi0_tm}) and (\ref{eq:tm_in_terms_of_chi})--(\ref{def:Lambda}) provide us with a complete knowledge of 
the electromagnetic scattering and subsequent excitation of graphene plasmons within the ribbons which make up the periodic system.

% *************************************************************
\subsection{Transmittance, reflectance and absorbance for normal incidence}

Here we consider the particular case where the impinging radiation strikes the graphene grid at normal incidence (see appendix 
\ref{appx:formulae} for oblique incidence), 
for which we have $k_x=0$ and $k_z= \sqrt{ \ep_{1}} k_0$, so that $q_n = nG$. In addition we remark that here, as in most 
experimental configurations, the $z$-component of the scattered wavevectors remains real only for the zero-th mode, 
i.e. $\kappa^{+/-}_{z,0} = \sqrt{ \ep_{2/1}} k_0$, while for the other diffraction orders it is imaginary, that is 
$\kappa_{z,n}^{+/-} = i\sqrt{ q^2_n - \ep_{2/1} k_0^2 }$. 
The reason for this is that often the period of the grating is much smaller than the 
impinging wavelength, $L \ll \lambda$, and thus $q^2_n \gg \ep_{2/1} k_0^2,\ \forall\ n \neq 0$. This is no coincidence, 
since our goal is to surpass the momentum imbalance between the incident light and GSPs. This can only be effectively achieved 
by fabricating subwavelength gratings. Therefore, we take $q_n > \sqrt{ \mathrm{max(}\ep_{1},\ep_{2} \mathrm{)} }  k_0$ for $n \neq 0$ henceforth. Consequently, 
only the zero-th mode reaches the far-field.

In possession of Eqs. (\ref{BC1:rm_Bi0_tm}) and (\ref{eq:tm_in_terms_of_chi})--(\ref{def:Lambda}), we have all the necessary ingredients 
to compute the scattering efficiencies for EM radiation striking the array of graphene ribbons. From the aforementioned 
expressions, the reflectance, transmittance and absorbance by the graphene grid read (see \ref{appx:formulae} for a detailed derivation)
\begin{align}
 R(\omega) &= \left| 1 - \frac{2\sqrt{\ep_1}}{\sqrt{\ep_1} + \sqrt{\ep_2}}  
 + \mu_0 \frac{\chi}{B_0^i} \frac{\pi w^2}{8 L} \frac{\sqrt{\ep_1}}{\sqrt{\ep_1} + \sqrt{\ep_2}} \right|^2 \ , \label{res:Refl_n0_theta0}\\
 T(\omega) &= \frac{ \Re\{1/\sqrt{\ep_2}\} }{ \Re\{1/\sqrt{\ep_1}\}} \left|\frac{\sqrt{\ep_2}}{\sqrt{\ep_1} + \sqrt{\ep_2}} 
 \left( 2  - \mu_0 \frac{\chi}{B_0^i} \frac{\pi w^2}{8 L} \right) \right|^2 \ , \label{res:Transm_n0_theta0}\\
 A(\omega) &= 1 - R(\omega) - T(\omega) \ , \label{res:Absor_n0_theta0}
\end{align}
respectively [and recall Eqs. (\ref{res:chi_result}) and (\ref{def:Lambda}) for $\chi$]. 
From the inspection of the above equations it is clear that the plasmonic resonances are controlled by the 
\emph{poles} of $\chi$ [or, similarly, $-\Im\mathrm{m} \left\{ \Lambda^{-1}(\omega) \right\}$]. Analyzing carefully the structure of the quantity $\Lambda(\omega)$, 
we readily identify that these occur whenever the condition
\begin{equation}
 \frac{\ep_1}{\sqrt{(nG)^2 - \ep_1 k_0^2}} + \frac{\ep_2}{\sqrt{(nG)^2 - \ep_2 k_0^2}} + i\frac{\sigma(\omega)}{\omega \ep_0} = 0
 \ , \label{eq:GSPs_condition}
\end{equation}
weighted by the factor $J_1(n \pi w / L)/n$ is met. Notice that Eq. (\ref{eq:GSPs_condition}) is nothing but the implicit expression for the 
dispersion relation of GSPs \cite{GoncalvesPeres,AbajoACSP,Xiao2016,ACSgp,Primer} with 
wavevector $q_n = n G$. However, what particular Bragg modes constitute the leading contributions 
for GSP-excitation strongly depends on the filling ratio $w/L$ (please refer to appendix \ref{appx:filling_ratio} 
for further details).

% ----------------------------------------------------------------------------
\section{Results and Discussion}

% *************************************************************
\subsection{Signatures of graphene plasmon resonances}

\begin{figure}[h]
  \centering
    \includegraphics[width=\the\columnwidth]{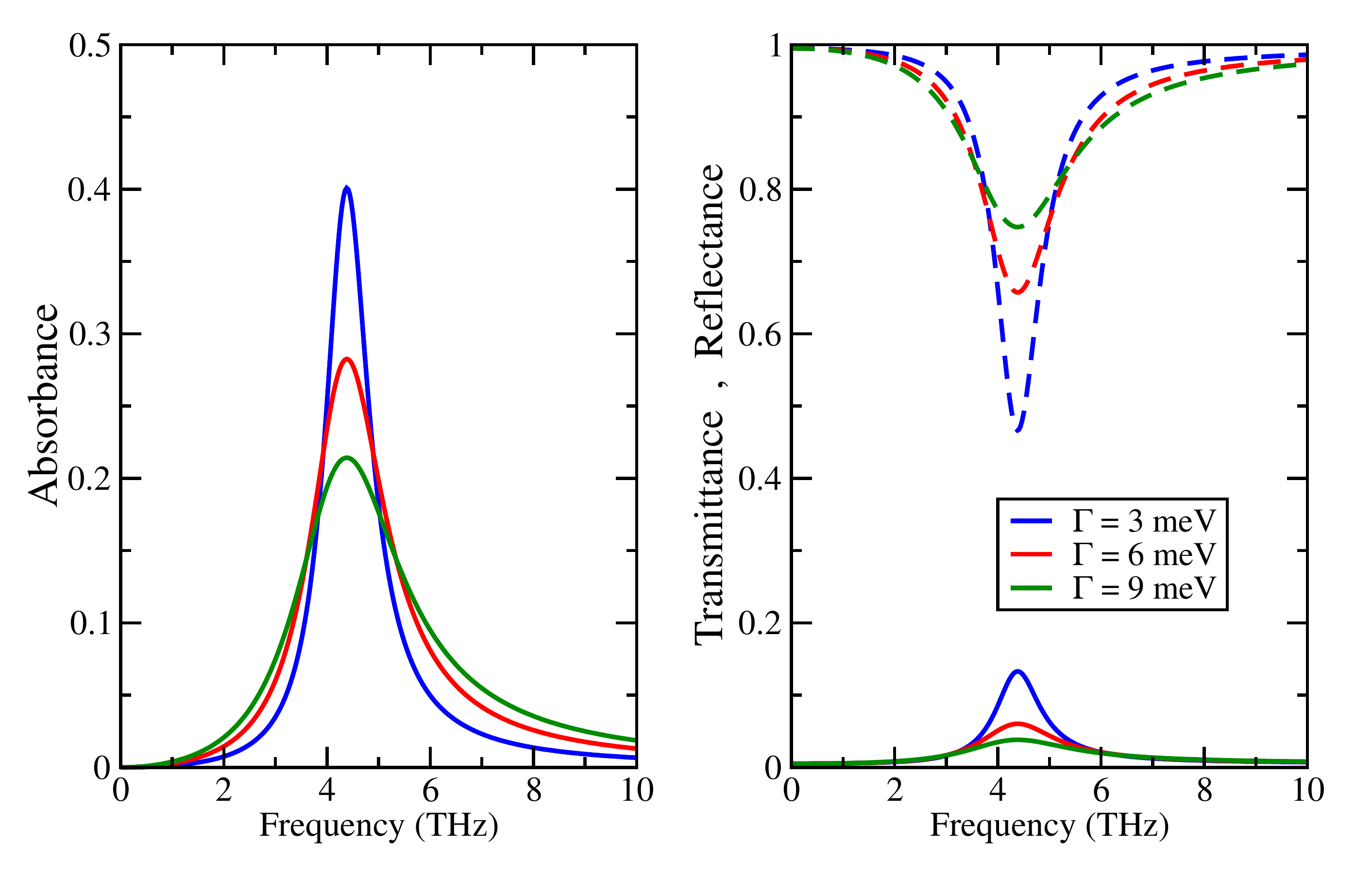}
      \caption[Absorbance, transmittance and reflectance spectra of a $p$-polarized plane-wave impinging on a periodic grid of graphene ribbons]
      {Absorbance spectra (left panel), transmittance and reflectance spectra (right panel) of a $p$-polarized plane-wave impinging on a periodic grid 
      of graphene ribbons for varying values of $\Gamma=\hbar\gamma$. 
      The remaining parameters are: $E_F=0.45$ eV, $w=2\ \mu$m, $L=4\ \mu$m, $\ep_1=3$, $\ep_2=4$, and $\theta=0$ (normal incidence).}
    \label{fig:G-grid_var_gamma}
\end{figure}

Having formulated our analytical model, we are now able to compute the absorbance, reflectance and transmittance spectra of EM radiation 
impinging on a periodic grid of graphene ribbons at normal incidence. Therefore, the results presented below are based on 
the outcome of Eqs. (\ref{res:Refl_n0_theta0})--(\ref{res:Absor_n0_theta0}). In the following, we take the conductivity of the graphene ribbons 
as the Drude conductivity of bulk-graphene (see appendix \ref{appx:sigma_g}). This is 
a rather good approximation for doped graphene ribbons in the THz spectral range, 
as long as the ribbons are not \emph{too small} (e.g. wider than several tens of nanometers \cite{ACS6,PRB90:TC}).
In particular, in Fig. \ref{fig:G-grid_var_gamma} we show the absorbance spectra (left panel) and corresponding reflectance and 
transmittance spectra (right panel) for different values of the damping parameter, $\Gamma$. 
The main feature figuring in the various 
spectra is the presence of a well-defined peak in absorption signalling the excitation of graphene plasmons. This GSP-assisted effect yields 
a dramatic enhancement in the absorbance spectra of the grating, owing to the coupling of free-propagating THz radiation to plasmons supported by 
the graphene ribbons which compose the periodic grid. Note that the aforementioned GSP-induced absoption comes hand in hand with a supression 
in transmittance and with an increase in reflectance at the GSP resonant frequencies, which roughly correspond to the poles of $\chi$ 
(or, in other words, the zeros of $\Lambda$).
Without surprise, smaller values of $\Gamma$ render sharper resonances, 
with these becoming successively broader and less pronounced as the electronic scattering rate increases. Also, 
the resonance frequency of the GSPs modes 
depends weakly on the value of $\Gamma$, as it remains essentially unchanged despite the different values of the damping parameter.

It should be stressed that, in principle, the interaction of EM radiation with the grating gives rise to multiple plasmon resonances 
with $q \approx (2m+1)\pi/w$ (for $m=0,1,2,...$).\cite{nphoton7,PhysRevB.71.035320,GoncalvesPeres} The most prominent 
resonance corresponds to the fundamental plasmon mode, while the 
higher-order resonances become increasingly weaker. Note that in the present model the latter are necessarily ignored, owing to the choice 
of ansatz for the current [cf. Eq. (\ref{eq:Jx_ansatz})] which can only account for the dipole-like fundamental resonance (the 
one that appears in Fig. \ref{fig:G-grid_var_gamma}). Fortunately, 
this resonance carries most of the spectral weight (see appendix \ref{appx:weight}) and it
clearly dominates the polaritonic spectrum;\cite{Primer,GoncalvesPeres}  
in fact, the resonances that emerge at higher frequencies are often invisible (or barely visible) in 
many experiments, since they can only be detected for small values of $\Gamma$.

\begin{figure}[h]
  \centering
    \includegraphics[width=\the\columnwidth]{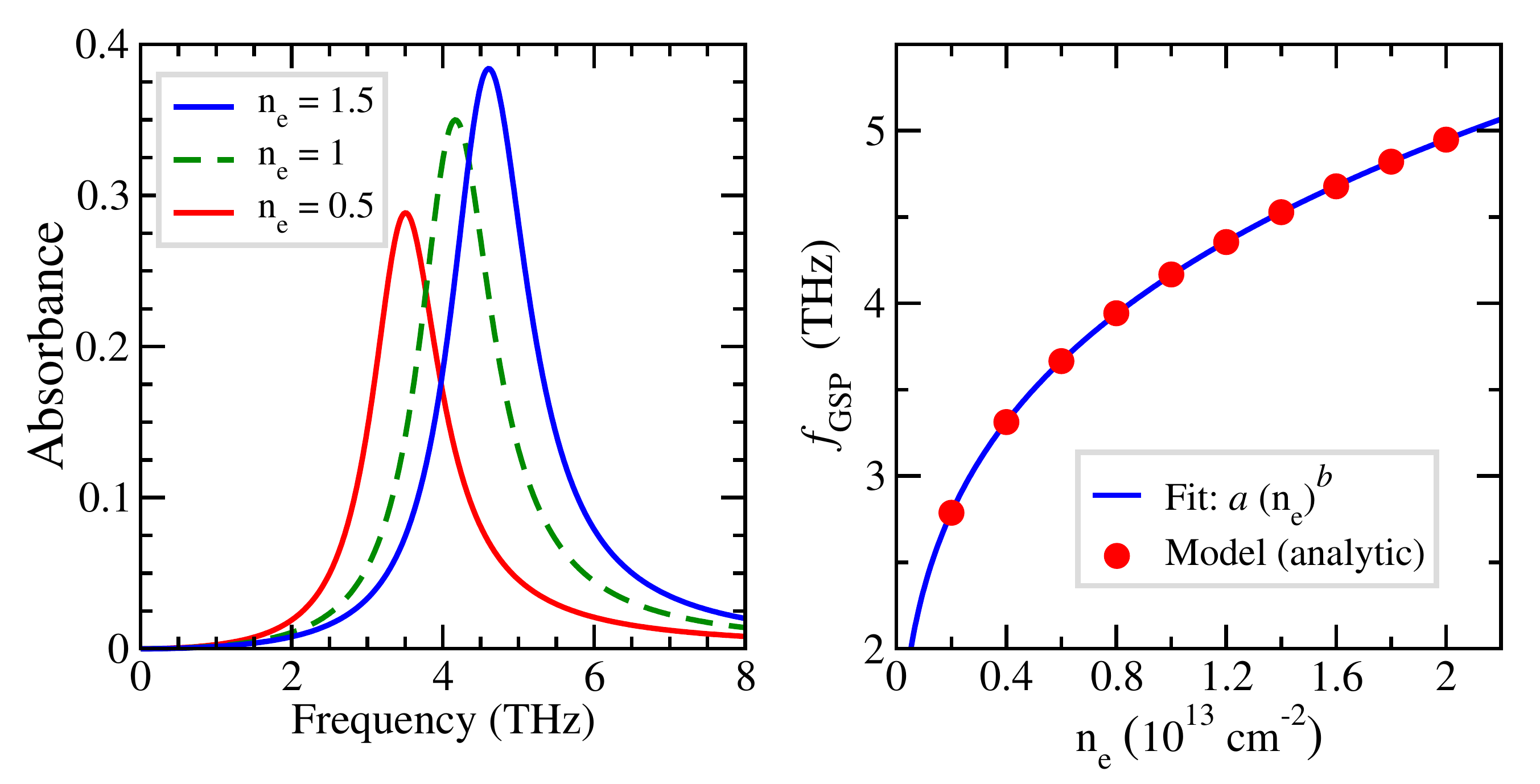}
      \caption[Dependence of the GSP-frequency on the electronic density in graphene]{Dependence of the GSP-frequency on the electronic density 
      of the graphene ribbons. Left panel: absorbance spectra for different selected values of $n_e$; the legend gives $n_e$ in units 
      of $10^{13}$ cm$^{-2}$. Right panel: resonant frequencies --- retrieved from our
      analytic theory (points) --- for several values of $n_e$, and corresponding fitting function to those points, that is 
      $f_{\mbox{\tiny GSP}}(n_e) \propto n_e^b$ with $b=0.249\simeq1/4$, in accordance with what is expected for graphene plasmons (see main text).
      Parameters: $\Gamma=3.7$ meV, $w=2\ \mu$m, $L=4\ \mu$m, $\ep_1=3$, $\ep_2=4$, and $\theta=0$ (normal incidence). We have used 
      $E_F = \hbar v_f \sqrt{\pi n_e}$, with $v_F \approx 1.1 \times 10^6$ m/s.}
    \label{fig:G-grid_var_ne}
\end{figure}

We now explore the dependence of the GSP-induced absorption spectra on the different parameters of the system. One of the most important 
parameters is the electronic density, $n_e$. This quantity is related with the material's Fermi energy via $E_F = \hbar v_F \sqrt{\pi n_e}$, where 
$v_F \approx 1.1 \times 10^6$ m/s is the Fermi velocity of the Dirac fermions in graphene.\cite{RevModPhys.84.1067,DiracIRspec}  
The density of graphene charge-carriers 
can be easily controlled by means of electrostatic gating. This possibility is of extreme relevance in graphene plasmonics, 
since it enables the 
excitation and control of tunable GSPs with tailored properties at the distance of a voltage knob. 
The effect of varying the electronic density within the graphene ribbons which 
constitute the grating is demonstrated in Fig. \ref{fig:G-grid_var_ne}. 
From the figure, it is clear that GSP-resonances become stronger and shift toward higher frequencies as the density 
of charge-carriers increases. In order to quantify such behavior, in the right panel of Fig. \ref{fig:G-grid_var_ne} we have plotted
the GSP-frequency (corresponding to the fundamental mode) as a function of the doping level, 
to which we have fitted a function of the type $f_{\mbox{\tiny GSP}}(n_e) \propto n_e^b$, having obtained $b=0.249\simeq1/4$ for the exponent 
(fitting parameter)\footnote{The fitting function was obtained using the least-squares method.}.
This therefore demonstrates that the observed resonances scale with the electronic density as 
\begin{equation}
 f_{\mbox{\tiny GSP}} \propto n_e^{1/4}%\ ,,
\end{equation} 
which 
is a specific signature of graphene plasmons.\cite{NatNano,nphoton7,Primer,GoncalvesPeres} Contrariwise, 
in typical 2DEGs a scaling with $n_e^{1/2}$ is observed instead. The 
different scaling for graphene is a direct consequence of the linear dispersion exhibited by the Dirac particles in this material. \cite{Geim09,RMP81} 

\begin{figure}[h]
  \centering
    \includegraphics[width=\the\columnwidth]{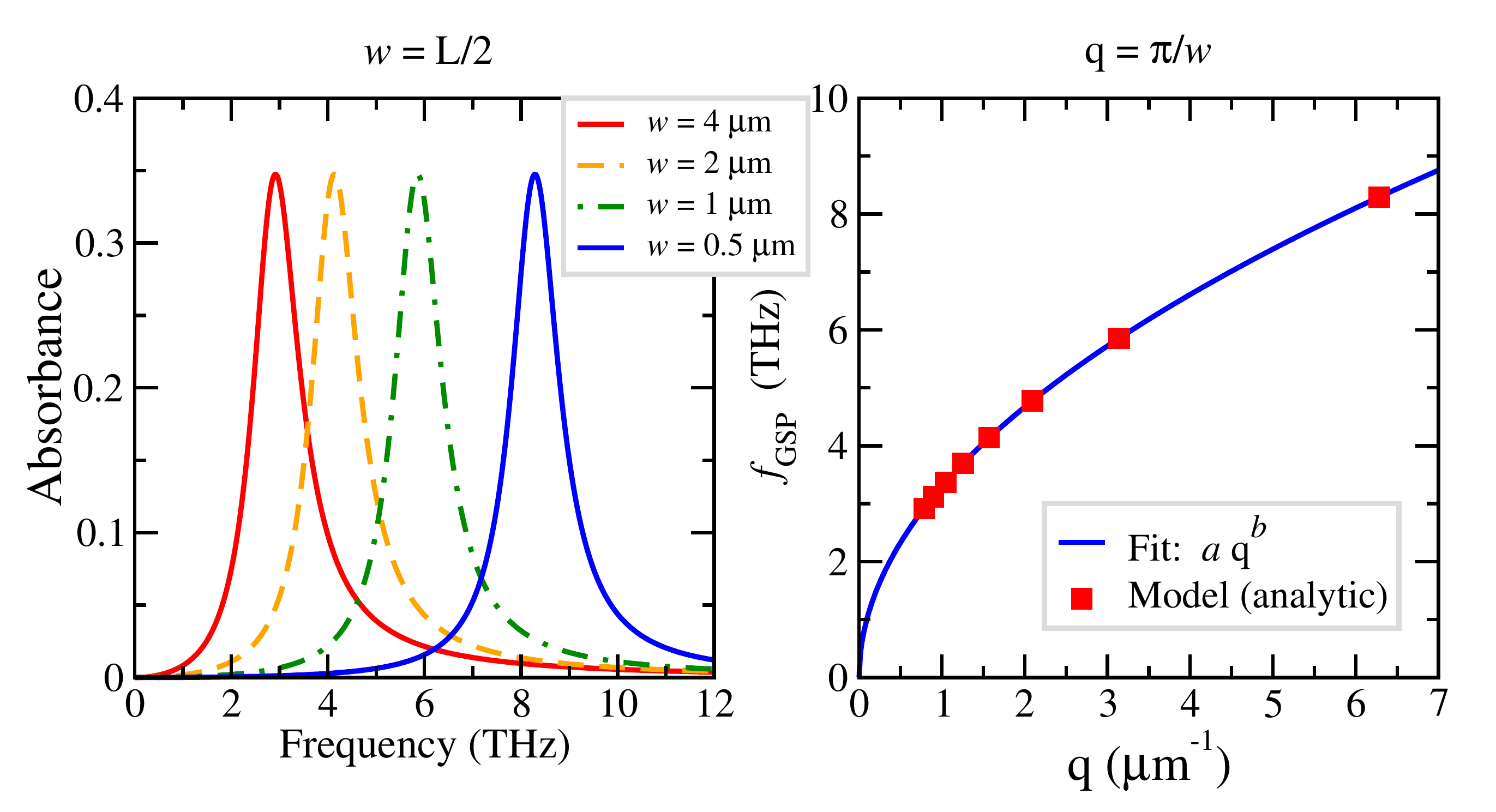}
      \caption[Dependence of the GSP-frequency on the period of the grid]{
      Dependence of the GSP-frequency on the period of the graphene grating, $L$. We kept the ratio $w=L/2$ fixed, and therefore 
      the GSP-fundamental mode carries a wavevector of about $q \sim \pi/w$. In the right 
      panel we show the position of the GSP-resonances rendered by the analytical method incorporating the edge condition as a 
      function of the GSP wavevector (square data points). To these data we have fitted a curve $f_{\mbox{\tiny GSP}}(q) \propto q^b$, having 
      obtained a exponent $b=0.502 \sim 1/2$. Parameters: $E_F=0.4$ eV, 
      $\Gamma=3.7$ meV, $\ep_1=3$, $\ep_2=4$, and $\theta=0$.}
    \label{fig:G-grid_var_q}
\end{figure}

An alternative way to tune the GSP-resonances is to pattern grids of graphene ribbons of different widths, $w$. For the sake of clarity, 
we shall keep the filling ratio $w/L=1/2$ constant. Figure \ref{fig:G-grid_var_q} depicts the calculated absorbance spectra 
for periodic arrays of graphene ribbons of different widths. Notice that structures with narrower ribbons yield GSPs 
with higher energies. 
The fundamental plasmonic resonance of the grating resembles the excitation of a GSP in extended 
graphene with $q \sim \pi/w$. Such large wavevectors can only be attained due to the contribution of the several Bragg 
diffraction orders originating from the interaction of the incident light with the grid. Within this reasoning, 
Fig. \ref{fig:G-grid_var_q} shows that the plasmonic resonances scale as $\sqrt{q}$, 
in a similar way that plasmons in an unpatterned, continuous graphene sheet do.
We have also found that, while the filling ratio $w/L$ has a significant impact on the position of the plasmonic 
resonance of the graphene grid, the physics of system is largely determined by the ribbon-size, i.e. $f_{\mbox{\tiny GSP}} \propto w^{-1/2}$
provided that $w/L < 1/2$, that is, that the interaction between neighboring ribbons is small.

\begin{figure*}%[h]
  \centering
    \includegraphics[width=0.85\textwidth]{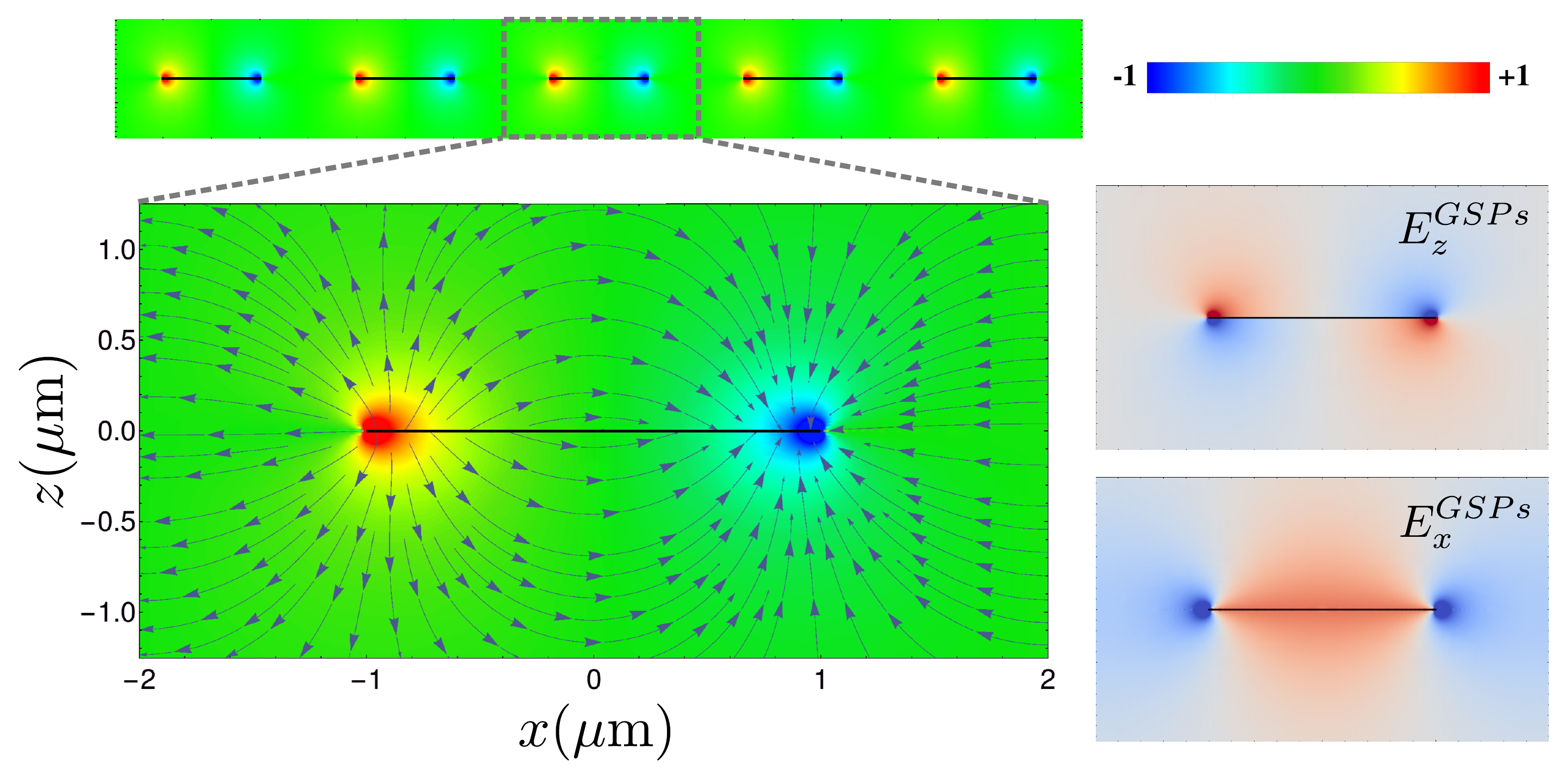}
      \caption[Electromagnetic field (normalized) akin to GSPs in the grating's unit cell]{
      Electric field representing graphene plasmons excited in the graphene ribbons which compose the grid 
      (with dimensions $w=2\ \mu$m and $L=4\ \mu$m). The figures show the 
      plasmonic fields in the system's unit cell, and the graphene ribbon is indicated by the horizontal black line. 
      Left panel: Vectorial representation of the electric field (in the $y=0$ plane) 
      due to GSPs, $\mathbf{E}^{GSPs}(x,z) = E_x^{GSPs}(x,z) \mathbf{\hat{x}} + E_z^{GSPs}(x,z) \mathbf{\hat{z}}$. The intensity 
      plot refers to the quantity $\sgn(z)E_z^{GSPs}$ which roughly highlights the charge-density within each graphene ribbon. 
      Righ panel: normalized spatial distributions of the electric field components $E_z^{GSPs}$ (top) and $E_x^{GSPs}$ (bottom). The spatial 
      range covered in these sub-panels is the same as in the main panel. We note that 
      only modes corresponding to plasmonic (evanescent) modes were included in the sums figuring 
      in the Bloch expansions, which for the parameters used here encompass all modes with the exception 
      of the specular one (i.e. with $n=0$). 
      Parameters: $E_F = 0.4$ eV, $\Gamma = 3.7$ meV and $\epsilon_1 = \epsilon_2 = 4$.}
    \label{fig:E_field}
\end{figure*}

The calculated electric field akin to the graphene plasmons supported by the ribbons which constitute the grating 
is depicted in Fig. \ref{fig:E_field} (depicting the unit cell of a representative 
array of ribbons with dimensions $w=2\ \mu$m and $L=4\ \mu$m). Such computation is straightforward once 
the resonant frequency akin to the fundamental GSP mode is determined; that information is then fed into the
Bloch amplitudes (\ref{BC1:rm_Bi0_tm}) and (\ref{eq:tm_in_terms_of_chi}) that represent the scattered field. %
The figure plainly demonstrates the potential graphene plasmons have to squeeze EM fields into deep 
subwavelength dimensions. Notice that most of the modal energy is concentrated in the immediate vicinity 
of the graphene ribbons. In addition, the spatial distribution of the electric field is not uniform 
along the ribbons' transverse direction: the density of charge-carriers (and thus the electric field) 
is higher at both edges of the ribbons. The charge-density is also antisymmetric with respect to the ribbons' 
midpoint, bearing some similarity to an electric dipole.  
Such extreme field localization plays a pivotal role, for instance, in biosensing, 
allowing the detection of minute variations in the local dielectric environment 
due to the presence/adsorption of a given target analyte. This property, together 
with ability to tune the GSP-resonances, enables not only unprecedentedly large field 
overlaps, but also provides a route to tailor the interaction of GSPs with the 
vibrational resonances of biochemical molecules, thereby achieving huge spectral 
overlaps that allow specific label-free detection of biomolecules via their vibrational 
fingerprints.\cite{Rodrigo10072015,ACSphot6b00143}

We further note that the overall spatial configuration of the field illustrated in Fig. \ref{fig:E_field} 
is qualitatively maintained throughout a wide range of ribbon widths, from the micrometer to the nanometer size, 
provided that we are within the quasi-static regime (i.e. $q_{\mbox{\tiny GSP}} \gg k_0$) and at resonant frequencies 
below $E_F/\hbar$, beyond which GSPs become quenched owing to the onset of interband Landau damping.\cite{AbajoACSP,GoncalvesPeres} 
This scale-invariance is a property of the electrostatic limit.\cite{ACS6Chris,AbajoACSP,GoncalvesPeres} Note, however, 
that we include retardation in our calculations nevertheless.

The results presented above, covering an appreciable vast parameter space, suggest that our analytical model 
is able to correctly describe the fundamental plasmonic excitations which arise in periodic grids of graphene 
ribbons. We thus have built an  analytic framework which delivered closed-form expressions 
for the spectra which can be easily evaluated, and that yield results consistent with those found in the 
literature.\cite{NatNano,nphoton7,Primer,Xiao2016,ACSgp,AbajoACSP,GoncalvesPeres}

% *************************************************************
\subsection{THz plasmons in graphene microribbons}

In order to determine to what extent our analytical model is capable of explaining 
experimental spectra, we shall test our theory against measured data taken 
from the experiments performed by Ju \emph{et al.}\cite{NatNano} To that end, 
we mimic the experimental setup by feeding the reported empirical parameters into our 
equations. In addition, we concurrently employ a semi-analytical technique 
introduced elsewhere\cite{JPCM24,PolCrysPRB85,JOpt15,Primer,GoncalvesPeres} 
which represents the conductivity of graphene (and resultant current) in 
terms of a Fourier series (see appendix \ref{appx:semi-ana}). This has the advantage of taking into account not only 
the fundamental plasmonic resonance, but also the higher-order ones. On the other hand, 
it requires the numerical solution of a linear algebra problem and therefore 
we refer to it as a \emph{semi-analytical} method hereafter. 
We further emphasize that, to the best of our knowledge, so far no attempt has been made 
to perform a direct comparison of the outcome of this latter method against available experimental data. 
Nevertheless, it is still far less computationally-demanding than fully numerical simulations 
such as the FDTD or FEM techniques.\cite{CEM} 

In their experiments, the authors of Ref. [\onlinecite{NatNano}] fabricated three different samples 
containing periodic grids of graphene ribbons with widths of 4, 2 and 1 $\mu$m, while 
maintaining the ratio $L=2w$ unchanged. Moreover, the authors have concluded that an effective 
dielectric constant of $\ep_{\mathrm{eff}}=5$ adequately accounts for the intricate optical constants 
of the cladding dielectrics (ion gel and SiO$_2$/Si), and have reported a scattering 
rate of $\gamma/(2\pi) = 4$ THz.\cite{NatNano} While at THz frequencies the conductivity of graphene 
can be approximated by its Drude expression, here we model the conductivity as 
obtained using the Kubo formula at room temperature (see appendix \ref{appx:sigma_g}). This is necessary here  
because the experimental data of Ju \emph{et al.}\cite{NatNano} refers to 
the change in transmittance with respect to the same quantity measured at the 
``charge neutral point'' (CNP) [when the Fermi level is at the so-called Dirac point] 
where neither finite-temperature nor interband processes can be neglected.

\begin{figure*}%[h]
  \centering
    \includegraphics[width=0.8\textwidth]{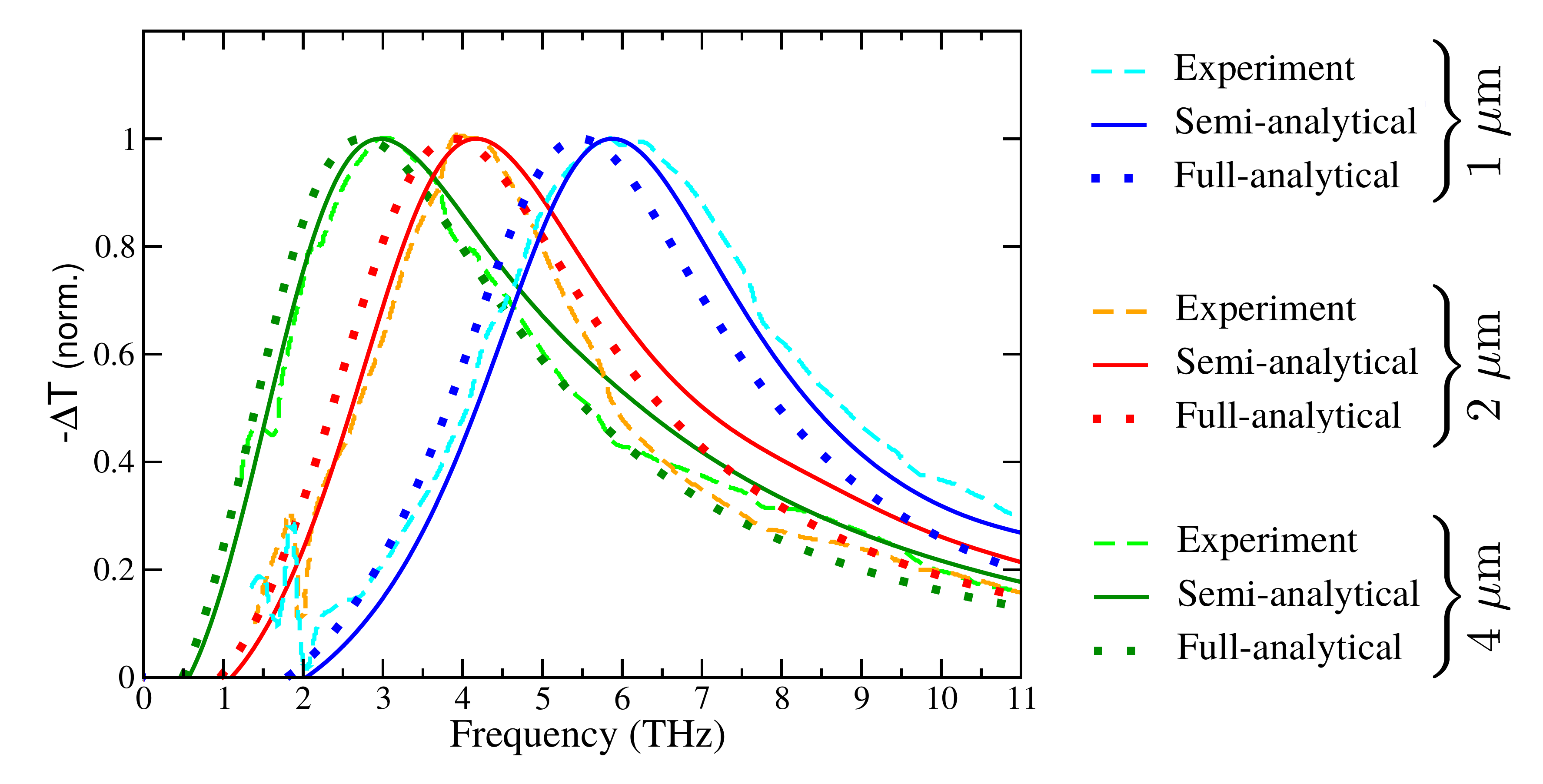}
      \caption[Comparison between theory and experimental data]{
      Normalized plasmon-induced change in transmittance relative to the CNP, $-\Delta T = T_{CNP} - T$, 
      in periodic grids of graphene ribbons with different dimensions: $w=4\ \mu$m (green), $w=2\ \mu$m (red), $w=1\ \mu$m (blue), 
      and $w/L=1/2$ throughout. The dashed lines correspond to experimentally measured spectra, \cite{NatNano} while 
      the dotted lines and solid lines correspond to the spectra obtained using our full-analytical model and 
      the semi-analytical technique (see appendix \ref{appx:semi-ana}), respectively. 
      We have used the following parameters, in accordance 
      with Ref. [\onlinecite{NatNano}]: $E_F=0.497$ eV, $\Gamma = 16.5$ meV and $\epsilon_1=\epsilon_2=5$.}
    \label{fig:Theo_vs_Exp}
\end{figure*}

The comparison between the calculated GSP-induced (normalized) change in transmittance, $-\Delta T = T_{\mathrm{CNP}} - T$, 
and the experimental data is portrayed in Fig. \ref{fig:Theo_vs_Exp}. The observed agreement 
between theory and experiment is outstanding and constitutes compelling evidence that 
both theories are capable of interpreting the measured spectra.
The peaks visible in the figure originate from the excitation of the main GSP plasmonic resonance 
supported by the graphene ribbons which form the grid. The shifting of the plasmon resonances 
towards higher frequencies as a consequence of the narrowing of the ribbons 
exhibits the predicted $f_{\mbox{\tiny GSP}} \propto w^{-1/2}$ scaling behavior. 
Although the degree of accordance between the data and the full-analytical model 
is rather good, it seems that this model slightly underestimates the 
resonant frequency. On the other hand, the agreement among the 
experimental measurements and the spectra obtained by the semi-analytical model 
is indeed quite remarkable, with the computed (measured) GSP-resonances 
located at 2.9 (3), 4 (4.1) and 5.6 (6) THz for arrays with ribbon widths 
of 4, 2 and 1 $\mu$m, respectively. Therefore, a possible motif for 
the small redshift visible in the spectra produced by the 
full-analytic theory may be concerned either with the fact that it neglects the 
higher-order plasmon resonances and/or to the approximation 
made for the current involving the edge condition, for instance, it may not be 
\emph{exactly} modeled by a square-root as in Eq. (\ref{eq:Jx_ansatz}) [see appendix \ref{appx:ansatz}]. Still, and 
despite this fact, the fidelity of the analytical model remains very good.

It should be appreciated that the use of either one of the above-mentioned analytical or 
semi-analytical techniques, apart from requiring less resources, 
provide a level of physical insight and intuition that numerical methodologies based on the numerical solution of 
Maxwell's equations simply cannot envision.

% *************************************************************
\subsection{Hybrid mid-IR plasmons in graphene nanoribbons: plasmon-phonon coupling}

Graphene plasmonics has the potential to become a viable tool for nanophotonic devices 
working within a broad spectral window, from the THz/far-IR up to mid-IR frequencies. 
Remarkably, routes to bring graphene plasmonics to near-IR and visible frequencies 
have already been proposed from a theoretical perspective\cite{AbajoACSP}.

\begin{figure*}%[h]
  \centering
    \includegraphics[width=0.85\textwidth]{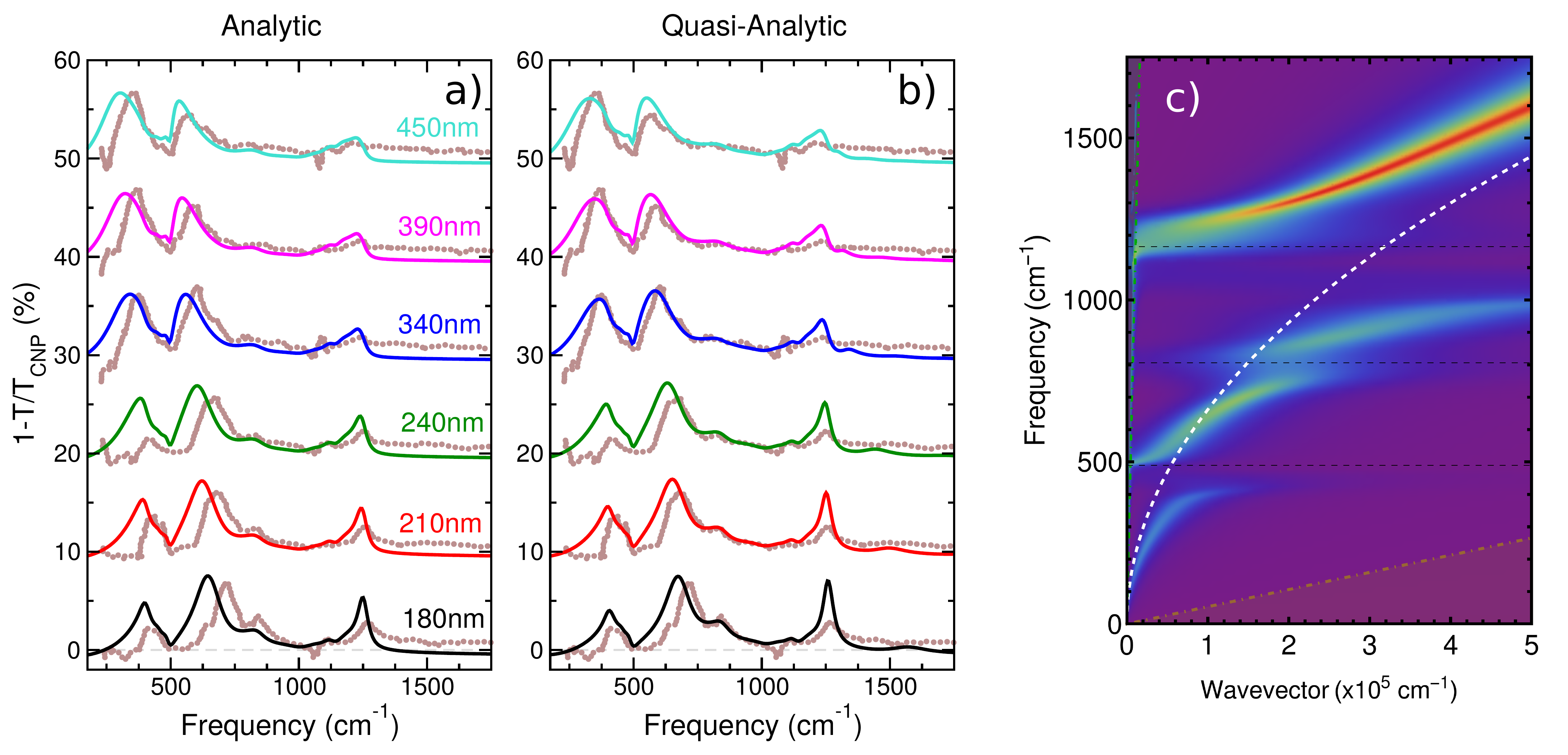}
      \caption[Hybrid mid-IR graphene plasmon-phonon polaritons]{
      Hybrid mid-IR graphene plasmon-phonon polaritons excited in period gratings of graphene nanoribbons 
      sitting on a polar (SiO$_2$) substrate. Experimental data (light-brown points) and corresponding extinction spectra 
      calculated using a) the analytical model; and b) the semi-analytical model (solid lines). The computations were 
      carried out in accordance with the experimental parameters\cite{luxmoore14}: $E_F=0.37$ eV, $\theta=0$ (normal incidence), 
      $\ep_1=1$ and the dielectric function of SiO$_2$, $\ep_2 \equiv \ep_{\mathrm{SiO}_2}(\omega)$, was taken 
      from the literature.\cite{Kitamura:07} We have used electronic scattering rates corresponding to about $25-30$ meV 
      depending on the sample. c) Loss function (via $\Im\mathrm{m}\ r_{\mathrm{TM}}$) for extended graphene deposited 
      on SiO$_2$, with the uncoupled GSP spectrum superimposed (white dashed line); we note that this serves only as 
      an eye-guide to the interpretation of the data, since we expect the polaritonic spectrum akin to 
      the periodic grid of nanoribbons to be slightly different from that of unpatterned, extended graphene.}
    \label{fig:Theo_vs_Exp_GSPPhPs_all}
\end{figure*}

At the time of writing, many experiments have reported GSPs at mid-IR 
frequencies.\cite{Rodrigo10072015,ACSphot6b00143,luxmoore14,ACS7,ZFnl14,nl13Brar,nlAtwater,nl14XZ}  
Coupling light to graphene plasmons at those frequencies can be realized in nanostructured graphene 
with typical dimensions from several tens of nanometers to a few hundreds of the nanometer. 
The mid-IR spectral range is a particularly important one, as many biological and chemical 
compounds exhibit resonances in that region of the EM spectrum. Thus, tunable graphene 
plasmons may be perceived as a fertile playground for applications in 
biochemical sensing and spectroscopy. Furthermore, when graphene is deposited 
in a polar substrate --- such as hBN or SiO$_2$ --- the Fuchs--Kliewer SO phonons\cite{PhysRev.140.A2076} 
of the substrate can couple to plasmons in graphene via Fröhlich interaction,\cite{Mahan}  
leading to the emergence of new hybrid modes dubbed 
graphene surface plasmon-phonon polaritons\cite{nphoton7,nlAtwater} (GSPPhs). In order to account 
for the optical phonons arising in the neighboring polar material(s), 
their corresponding frequency-dependent, 
complex-valued dielectric function(s) can be modeled using 
adequate Lorentz oscillator models\cite{AM} incorporating the phononic resonances, 
or more evolved models, e.g. based on Gaussian functions and integrals.\cite{Kitamura:07} 
The hallmark of strong coupling between graphene plasmons and SO phonons 
is the complete reshaping of the traditional $f_{\mbox{\tiny GSP}} \propto \sqrt{q}$ 
dispersion of bare GSPs into a set of multiple well-defined branches 
ascribed to hybrid GSPPhs modes possessing mixed plasmonic and phononic character, 
as demonstrated in Fig. \ref{fig:Theo_vs_Exp_GSPPhPs_all}-c) for extended graphene 
sitting on SiO$_2$ (see caption for further details). In particular, notice the evident anti-crossing 
behavior of the plasmon-phonon bands in the vicinity of the SO frequencies.

In Fig. \ref{fig:Theo_vs_Exp_GSPPhPs_all} [panels a) and b)] we compare 
the results for arrays of graphene ribbons obtained using 
the analytic [a)] and semi-analytic [b)] methods (solid lines)
against the experimental spectra (light-brown points) 
collected by Luxmoore \emph{et al.}\cite{luxmoore14} Their data 
shows evidence of strong interaction of GSPs --- excited in 
periodic grids of graphene nanoribbons --- with the three SO phonons 
of the underlying SiO$_2$ substrate. In our modeling, we have 
met the experimental configuration of the 
fabricated devices, consisting in doped ($E_F=0.37$ eV) nanoribbons with 
widths ranging from 450 nm down to 180 nm arranged in a periodic array with 
periodicity $L=5w/2$.\cite{luxmoore14} The dynamical dielectric function of the SiO$_2$ 
substrate was taken from the literature\cite{Kitamura:07}, and 
we have employed, as before, the optical conductivity of graphene under Kubo's framework.

It is worth to highlight that both the analytic and semi-analytic theories 
outlined above fit admirably to the experimental data, whose structure 
is now much more intricate than the one seen for microribbons in the THz range. 
The GSPPh-induced extinction spectrum ($1-T/T_{\mathrm{CNP}}$) of the several samples presented 
in Fig. \ref{fig:Theo_vs_Exp_GSPPhPs_all} reveals the existence of 
multiple peaks, which correspond to the four polaritonic bands visible in 
figure's last panel. We stress that, as expected, all four resonances shift toward higher 
frequencies upon decreasing ribbon size. Note, however, that they disperse 
at different rates. This is a direct consequence of the relative plasmon-to-phonon content 
which tends to vary depending on the distance each resonance is from the SO flat bands: 
the more plasmon-like the hybrid GSPPhs modes are (i.e. the farther they are 
from the uncoupled SO bands), the faster they disperse. Another key element, 
recognizable in the spectra, is the clear transfer of spectral 
weight from the first peak to the other resonances ascribed to higher GSPPhs bands. 
Together, the above-mentioned features constitute unambiguous manifestations of 
anti-crossing behavior. 

From the first two panels of Fig. \ref{fig:Theo_vs_Exp_GSPPhPs_all} 
it is apparent that both models closely follow the experimental data, 
thereby confirming their adequacy to describe the empirical extinction spectra. 
As in the case of the microribbons, the fully analytical method tends to 
underestimate slightly the position of the resonances, while the semi-analytical 
method yields an excellent agreement, particularly for the samples with wider ribbons. 
The small deviation from the data observed in the spectra of the narrower ribbons 
may have multiple origins exogenous to our theory, for instance, an incomplete 
knowledge of the dielectric properties of the \emph{particular} SiO$_2$ substrate used 
in the experiments and/or the effect of edge damage (and defects) introduced during 
the etching process\cite{nphoton7,luxmoore14,PRL98.206805} --- 
which, naturally, should be more pronounced for smaller ribbons ---, and 
may yield ribbon edges with impaired electrical activity (therefore 
rendering \emph{effective} widths smaller than the \emph{actual} ribbon widths\cite{nphoton7}). Here, we 
neglect the impact of the latter since it has been shown that the damage at the edges of each 
ribbon is highly heterogeneous.\cite{luxmoore14,PRL98.206805} 

The theoretical results produced in this work and subsequent confrontation against experimental data 
confirm the ability of the theoretical tools developed here to simulate and interpret 
spectra taken from real-world experiments, providing excellent, reliable results almost 
instantaneously\footnote{Using a standard personal laptop equipped with a dual-core (4 threads) $2.30$ GHz processor.} 
without the need of substantial computational resources.

% ================================================== // ==================================================

% *************************************************************
% ::                      CONCLUSIONS                        ::
% *************************************************************
\section{Conclusions}

In conclusion, we have developed a novel analytical approach, based on the edge condition 
and Bloch-expansions for the fields, to describe graphene plasmons excited in periodic grids of graphene ribbons. 
We solved the scattering problem and provided simple closed-form expressions to compute the reflectance, absorbance, transmittance, 
and related extinction spectra. We then benchmarked the results of our analytical 
theory using a semi-analytical model, and tested both techniques against experimental data 
available in the literature.\cite{NatNano,luxmoore14} Our results show a very good agreement 
between the theoretical curves and the empirical data, which constitutes compelling evidence 
for the validity of the aforementioned theories. That concordance extends 
from the THz, using microribbon arrays, to the mid-IR spectral region, using nanoribbons. 
In the latter domain, we have also investigated hybrid GSPPhs excitations that arise 
from the interaction of GSPs with the SO phonons of the SiO$_2$ substrate, 
leading to the appearance of four composite modes featuring spectral weight transfer, 
which is indicative of anti-crossing behavior (resulting from the reconstruction 
of the bare GSPs spectrum owing the polar coupling).

The approaches developed in this work have two main advantages: 
\begin{enumerate*}[label=(\roman*)]
 \item they endow us with a deeper insight and sense for the 
physics governing plasmonic excitations in engineered graphene structures; and,
 \item render viable simulations of experimentally relevant quantities, on-demand and almost instantaneously, 
 without the cost of lengthy, computationally-demanding full-wave numerical packages, at 
least for patterned structures with a fair degree of symmetry.
\end{enumerate*} 
On the other hand, these naturally cannot compete with fully-numerical techniques such as FEM simulations 
in what concerns versatility to deal with many different and complex geometries.

Our findings suggest that both the analytical and the semi-analytical models described here 
could be used to architecture new forefront nanophotonic experiments 
based on graphene plasmonics, which is emerging as a promising field to deliver 
cutting-edge optoelectronic devices with tailored light-matter interactions.

% ================================================== // ==================================================
% -------------------------------------------------------------
% #                      Acknowledgments                      #
% -------------------------------------------------------------

\begin{acknowledgments}
The authors thank N. Asger Mortensen for insightful and valuable comments.
PADG acknowledges financial support from Funda\c{c}\~{a}o para a Ci\^{e}ncia e a Tecnologia (Portugal) from grant No. PD/BI/114376/2016. 
NMRP and YVB acknowledge financial support from the European Commission through the project
 ``Graphene-Driven Revolutions in ICT and Beyond'' (Ref. No. 696656). 
 This work was partially supported by the Portuguese Foundation for Science and Technology 
 (FCT) in the framework of the Strategic Financing UID/FIS/04650/2013. The Center for Nanostructured Graphene is sponsored by the Danish National 
 Research Foundation, Project DNRF103.
\end{acknowledgments}

% -------------------------------------------------------------
% #                         Appendixes                        #
% -------------------------------------------------------------
\appendix

%\section{Further details and computations}
\section{}

\subsection{Convergence of the sum in $\Lambda(\omega)$}
\label{appx:sum}

Notice that our results for the reflection and transmission amplitudes fundamentally depend on $\chi$, which in turn strongly depends on 
the function $\Lambda(\omega)$. The latter reads [cf. Eq. (\ref{def:Lambda})]
\begin{align}
 \Lambda(\omega) &= \frac{w}{4} \sum_{n=-\infty}^{\infty} \frac{1}{n} J_1(n\pi w /L) 
 \left[ 1 + \frac{\sigma(\omega)}{\omega \ep_0} \frac{ \kappa^{+}_{z,n} \kappa^{-}_{z,n}}{\ep_1 \kappa^{+}_{z,n} + \ep_2 \kappa^{-}_{z,n}} \right]\nonumber\\
 &= \frac{\pi w^2}{8L} \left[ 1 + \frac{\sigma(\omega)}{\omega \ep_0} \frac{ \kappa^{+}_{z,0} \kappa^{-}_{z,0}}{\ep_1 \kappa^{+}_{z,0} + \ep_2 \kappa^{-}_{z,0}} \right]\nonumber\\ 
 &+ \frac{w}{2} \sum_{n=1}^{N} \frac{1}{n} J_1(n\pi w /L) 
 \left[ 1 + \frac{\sigma(\omega)}{\omega \ep_0} \frac{ \kappa^{+}_{z,n} \kappa^{-}_{z,n}}{\ep_1 \kappa^{+}_{z,n} + \ep_2 \kappa^{-}_{z,n}} \right]
 \ , \label{def:Lambda_appx}
\end{align}
where in the last equality we have made explicit use of the fact that the summand is even with respect to $n$, where $n \in$ integers. 
Note that this expression comprises an infinite sum over $n$, so that we have also truncated the sum in the last step of Eq. (\ref{def:Lambda_appx}) 
for numerical purposes. The question that now arises is: \textit{how large should $N$ be? And what requirements should it fulfill?}

\begin{figure}[h]
\centering
 \includegraphics[width=0.35\textwidth]{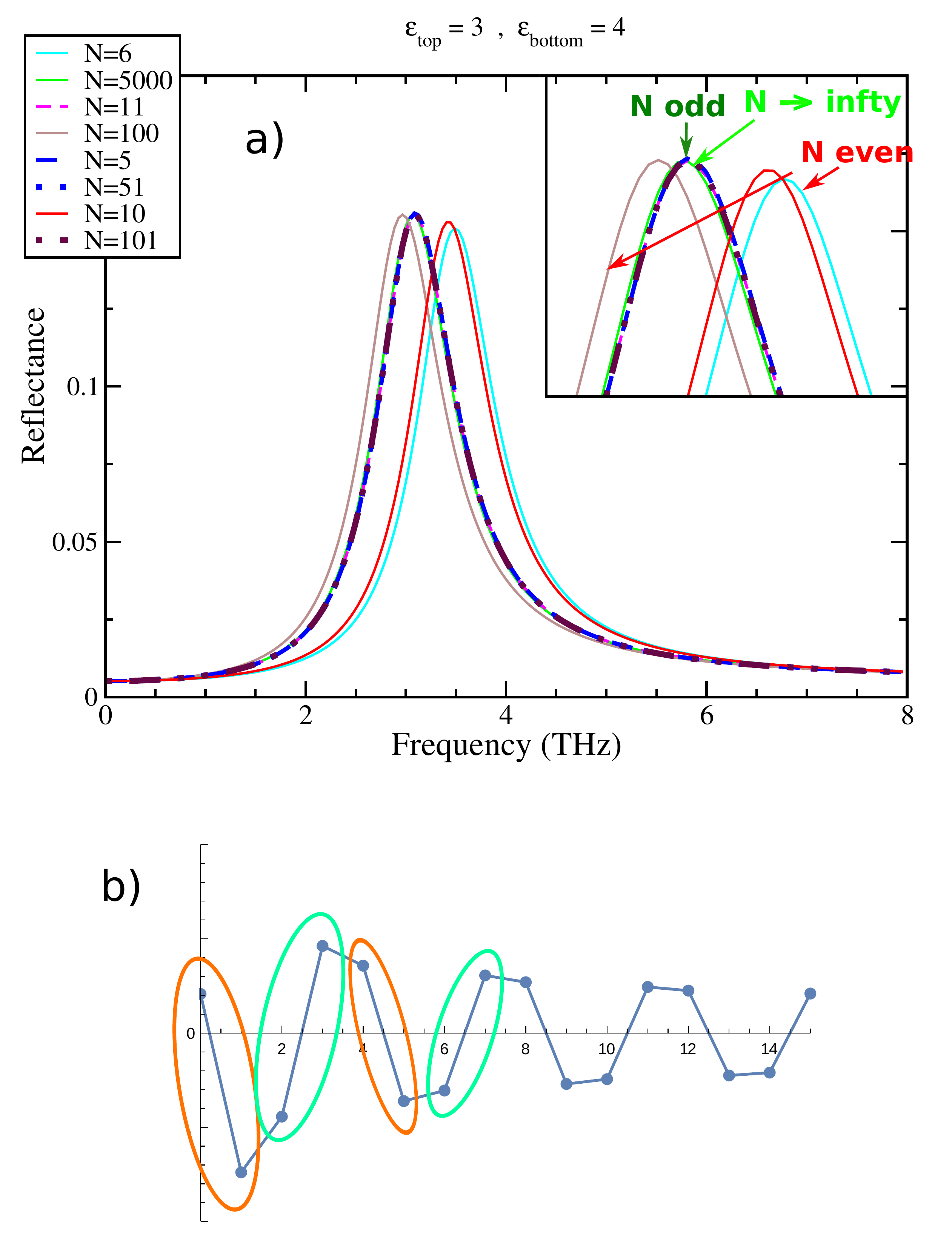}
  \caption[Results using different values for $N$]{
  a) Reflectance spectra using different values of $N$, which truncates the sum entering in the function $\Lambda(\omega)$. Notice the difference 
  between even and odd $N$-values, in the convergence of results.
  b) Alternating series figuring in $\Lambda(\omega)$ [cf. Eq. (\ref{def:Lambda_appx})] as a function of $n$.}\label{fig:N_conv}
\end{figure}

In order to answer these questions, let us plot the results of, say, the reflectance, using several values for $N$. The outcome of such procedure 
is shown in Fig. \ref{fig:N_conv}. From the figure, one can see a striking difference between the results obtained using odd $N$-values and 
even $N$-values. In particular, note that whenever we picked $N$ as odd we always got the same (converged) result. This result also coincides with the one in 
the limit $N \rightarrow \infty$ (light-green curve). Conversely, for even $N$-values, one sees an erratic behavior in the resulting spectra which 
indicates that the results did not converge. Only for very large values of $N$ even (such has 5000), one obtains the correct result.
Furthermore, notice that even when choosing a small, but odd value for $N$ -- such as $N=5$ --- one gets the same (correct) result that one would obtain by 
choosing a large, but even $N$ instead --- such as $N=5000$. This clearly highlights the need to choose $N$ correctly, namely to choose an odd-valued $N$.

The reason for this apparently counterintuitive behavior can be elucidated by plotting the values of the summand in Eq. (\ref{def:Lambda_appx}) 
as a function of $n$. This is done in panel b) of Fig. \ref{fig:N_conv}. Clearly, the figure shows an alternating 
series; this kind of series usually demands proper care for their accurate computation. In our case, one needs to have special care when choosing $N$, 
that is, to choose an odd-valued $N$. In that way, the total number of elements of the series is $N+1$ (because it includes the $n=0$ term), which is 
even, and, therefore, it correctly includes pairs of positive and negative values [as indicated in figure's \ref{fig:N_conv} panel b)].
Naturally, as $N$ approaches infinity, the choice between $N$-odd or $N$-even becomes unimportant. However, our analysis demonstrates that by using 
an odd-valued $N$, one can obtain accurate results with (odd) $N$-values as small as 5 (this truncation depends naturally on the particular 
system's geometry, but in general only a few terms of the sum are needed).

\subsection{Derivation of the formulae for the reflectance and transmittance spectra}
\label{appx:formulae}

For the derivation of the reflectance and transmittance scattering probabilities, we need to introduce the Poynting vector (for 
non-magnetic media),
\begin{equation}
 \mathbf{S} = \frac{1}{\mu_0} \mathbf{E} \times \mathbf{B}\ ,
\end{equation}
which characterizes the flux of electromagnetic energy per unit area. Assuming that both $\mathbf{E}$ and $\mathbf{B}$ 
can be written as in terms of harmonic functions, $e^{-i\omega t}$, one may write the time-averaged Poynting vector as
\begin{equation}
 \langle \mathbf{S} \rangle = \frac{1}{2\mu_0} \Re\mathrm{e}\{ \mathbf{E} \times \mathbf{B}^* \} \ ,
\end{equation}
where the star denotes the complex-conjugate. For a TM wave polarized in the $xz$-plane, the time-averaged Poynting vector reads
\begin{equation}
 \langle \mathbf{S} \rangle = \frac{1}{2\mu_0} \Re\mathrm{e}\{ E_x B_y^* \mathbf{\hat{z}} - E_z B_y^* \mathbf{\hat{x}} \} \ .
\end{equation}
Then, applying the previous equation for the incident wave, one obtains
\begin{equation}
 \langle S_z^i \rangle = \frac{c^2}{2\mu_0 \omega} \Re\mathrm{e} \left\{ \frac{k_z}{\ep_1} \right\} \left| B_0^i \right|^2 \ ,
\end{equation}
for the $z$-component (i.e. the component normal to the graphene grating). Similarly, for the reflected wave one has
\begin{equation}
 \langle S^r_{z,n} \rangle = -\frac{c^2 }{2\mu_0 \omega} \Re\mathrm{e} \left\{ \frac{\kappa_{z,n}^-}{\ep_1} \right\} \left| r_n \right|^2 \ , \label{Szn_r}
\end{equation}
for the $n$-th diffraction order (or Bloch-mode), whereas for the transmitted wave we obtain
\begin{equation}
 \langle S^t_{z,n} \rangle = \frac{c^2 }{2\mu_0 \omega} \Re\mathrm{e} \left\{ \frac{ \kappa_{z,n}^+ }{\ep_2} \right\} \left| t_n \right|^2 \ . \label{Szn_t}
\end{equation}
At this point we should stress that for purely imaginary wavevectors, $\kappa_{z,n}^{+/-} \rightarrow i\left|\kappa_{z,n}^{+/-}\right|$, the 
Poynting vectors associated with those Bloch modes give zero contribution [since $\Re\mathrm{e}\{ i|\kappa_{z,n}^{+/-}| \}=0$;
cf. Eqs. (\ref{Szn_r}) and (\ref{Szn_t})], and, as such, 
they will not contribute neither to the reflectance 
nor to the transmittance. This is because they are evanescent waves, and therefore they do not carry energy along the $z$-direction. 
Notice that the modes are so-called non-propagating or evanescent whenever $q_n > \ep k_0$, where $q_n = k_x + nG$ 
with $G=2\pi/L$, and $k_0=\omega/c$, in which case we have
\begin{equation}
 \kappa_{z,n}^{+/-} = \sqrt{\ep_{2/1} k_0^2 - q^2_n} \rightarrow i\sqrt{ q^2_n - \ep_{2/1} k_0^2 }\ ,
\end{equation}
in accordance with the definitions used in the main text.
Finally, the reflectance and transmittance through the structure under oblique incidence read
\begin{equation}
 R(\omega,\theta) = \sum_n \left| \frac{ \langle S^r_{z,n} \rangle }{ \langle S_z^i \rangle } \right|
 = \frac{ \Re\mathrm{e} \left\{ \kappa_{z,0}^- / \ep_1 \right\} }{ \Re\mathrm{e} \left\{ k_z/\ep_1 \right\} } \left| \frac{r_0}{B_0^i} \right|^2\ , \label{eq:Rn_appx}
\end{equation}
and
\begin{equation}
 T(\omega,\theta) = \sum_n \left| \frac{ \langle S^t_{z,n} \rangle }{ \langle S_z^i \rangle } \right|
 = \frac{ \Re\mathrm{e} \left\{ \kappa_{z,0}^+ / \ep_2 \right\} }{ \Re\mathrm{e} \left\{ k_z/\ep_1 \right\} }
 \left| \frac{t_0}{B_0^i} \right|^2\ , \label{eq:Tn_appx}
\end{equation}
respectively, where the sums were carried out over propagating modes only (solely the $n=0$ mode for the parameters used in this work). The absorption 
spectrum stems from these equations according to $A(\omega,\theta) = 1 - R(\omega,\theta) - T(\omega,\theta)$.\hfill\break

For normal incidence, the above formulae simplify considerably to:
\begin{align}
 R(\omega) &= \left| \frac{r_0}{B_0^i} \right|^2\ , \label{eq:R0_appx}\\
 T(\omega) &= \frac{ \Re\mathrm{e} \left\{ 1 / \sqrt{\ep_2} \right\} }{ \Re\mathrm{e} \left\{ 1 / \sqrt{\ep_1} \right\} }
 \left| \frac{t_0}{B_0^i} \right|^2\ , \label{eq:T0_appx}\\
 A(\omega) &= 1 -\left| \frac{r_0}{B_0^i} \right|^2 
 - \frac{ \Re\mathrm{e} \left\{ 1 / \sqrt{\ep_2} \right\} }{ \Re\mathrm{e} \left\{ 1 / \sqrt{\ep_1} \right\} }
 \left| \frac{t_0}{B_0^i} \right|^2\ ,
\end{align}
where $r_0$ and $t_0$ are computed using Eqs. (\ref{BC1:rm_Bi0_tm}) and (\ref{eq:tm_in_terms_of_chi})--(\ref{def:Lambda}). Explicitly, 
\begin{align}
 \frac{t_0}{B_0^i} &= \frac{\sqrt{\ep_2}}{\sqrt{\ep_1} + \sqrt{\ep_2}} 
 \left[ 2  - \mu_0 \frac{\chi}{B_0^i} \frac{\pi w^2}{8 L} \right]\ , \\
 \frac{r_0}{B_0^i} &= 1 - \frac{2\sqrt{\ep_1}}{\sqrt{\ep_1} + \sqrt{\ep_2}}  
 + \mu_0 \frac{\chi}{B_0^i} \frac{\pi w^2}{8 L} \frac{\sqrt{\ep_1}}{\sqrt{\ep_1} + \sqrt{\ep_2}} \ , 
\end{align}
with
\begin{align}
 \frac{\chi}{B_0^i} &= \frac{2 \kappa^{+}_{z,0} \kappa^{-}_{z,0}}{\ep_1 \kappa^{+}_{z,0} + \ep_2 \kappa^{-}_{z,0}} 
 \frac{\sigma(\omega) c^2}{\omega} \frac{1}{\Lambda(\omega)}
\ , \\
 \Lambda(\omega) &= \frac{w}{4} \sum_{n=-\infty}^{\infty} \frac{1}{n} J_1(n\pi w /L) 
 \left[ 1 + \frac{\sigma(\omega)}{\omega \ep_0} \frac{ \kappa^{+}_{z,n} \kappa^{-}_{z,n}}{\ep_1 \kappa^{+}_{z,n} + \ep_2 \kappa^{-}_{z,n}} \right]
 \ . 
\end{align}

\newpage

\subsection{Modes contributing to the fundamental resonance for different filling ratios}
\label{appx:filling_ratio}

We further note that the particular Bragg modes that 
couple the most to the fundamental GSP-resonance depend on the specific filling ratio of the system. Such dependence is 
a consequence of the overlapping of the fields with the ribbon, and, therefore, the geometric configuration (the filling ratio) is an 
important parameter. 
As an example, in Fig. \ref{SIfig:n_res} (obtained using the 
semi-analytical method outlined in appendix \ref{appx:semi-ana}) we shown relative modulus-squared 
Bloch amplitudes $| E_{x,n}^{(2)} |^2$ corresponding to several Bragg modes, for two different configurations.
\begin{figure}[h!]
  \centering
    \includegraphics[width=0.4\textwidth]{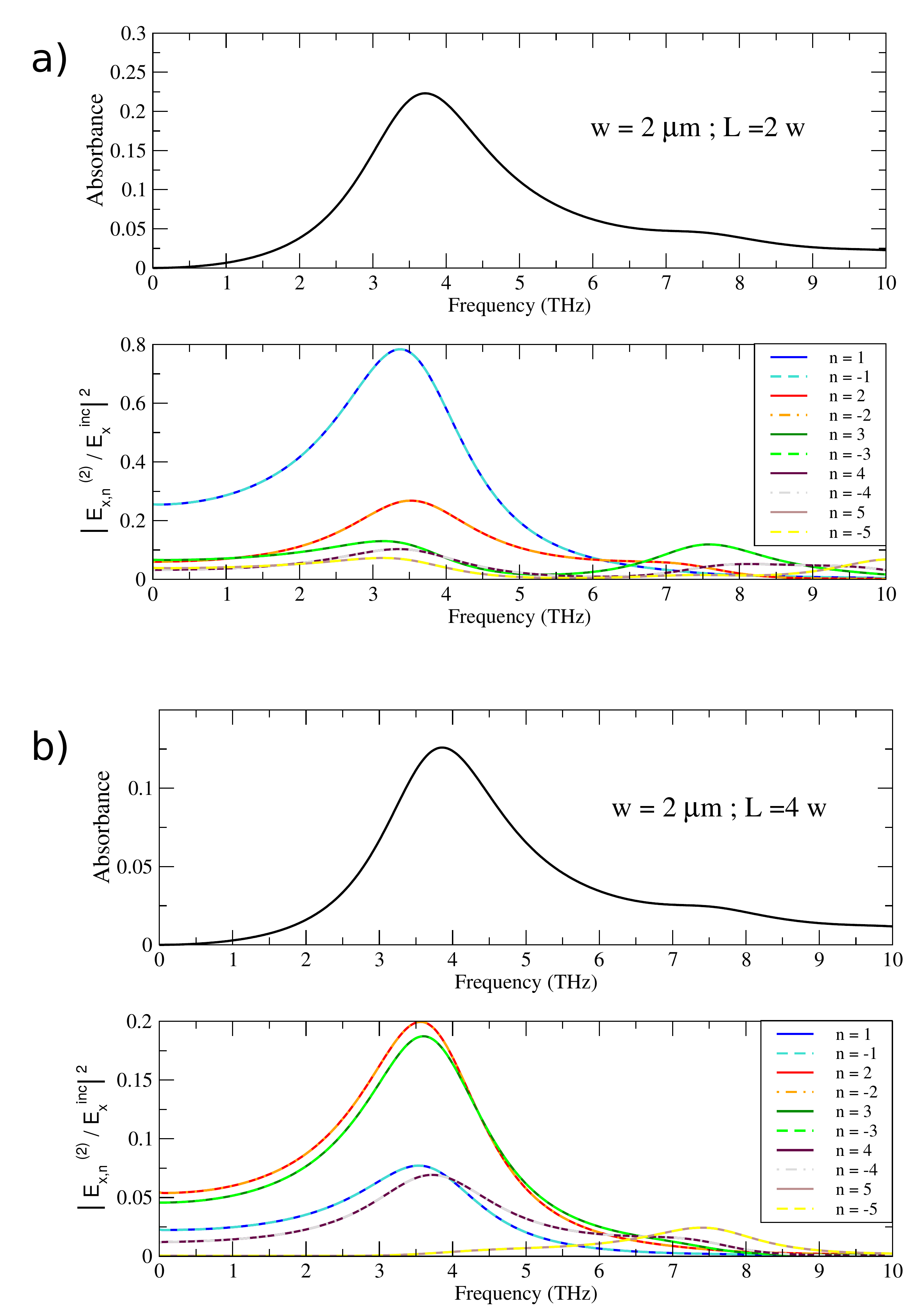}
      \caption{Absorbance and relative modulus-squared Bloch amplitudes $| E_{x,n}^{(2)} |^2$ corresponding to several Bragg modes, for two different 
      configurations: one with a filling ratio of $w/L=1/2$ (top) and another with $w/L=1/4$ (bottom).}
    \label{SIfig:n_res}
\end{figure}

\subsection{Spectral weight akin to the higher-order multipolar resonances}
\label{appx:weight}

\begin{figure}[h!]
  \centering
    \includegraphics[width=0.4\textwidth]{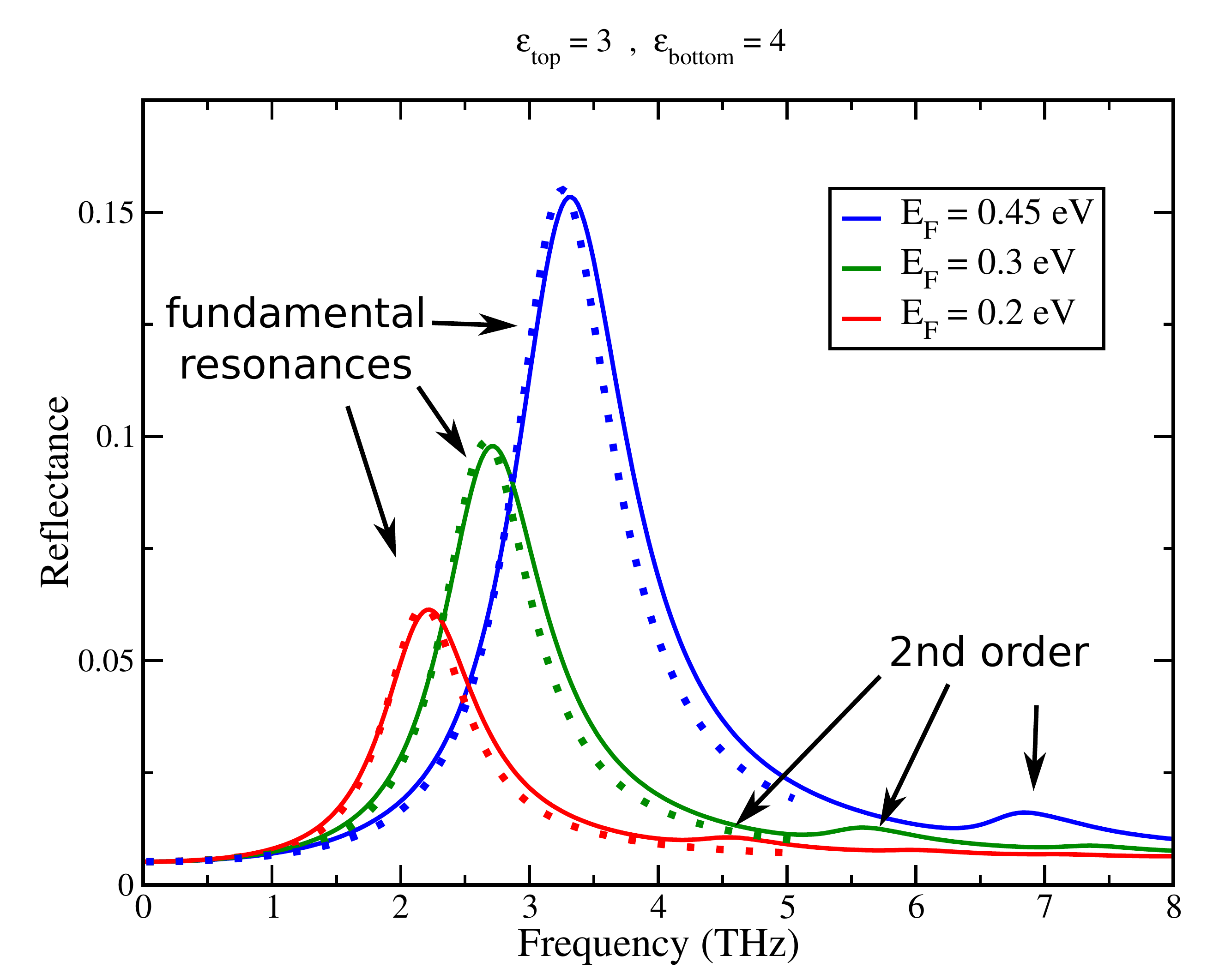}
      \caption{GSP-resonances in a periodic grid of graphene ribbons ($w=4\ \mu$m and $L=2w$) at different carrier concentrations. 
      The solid lines correspond 
      to the results obtained using the semi-analytical method whereas the dotted lines correspond to the outcome of the full-analytical approach.}
    \label{SIfig:spectralWeight}
\end{figure}

Figure \ref{SIfig:spectralWeight} shows the two lowest-energy plasmonic resonances in the system under study (for different 
doping levels). 
Note that, as discussed in the main article, the analytic approach does not account for the higher-order (multi-polar) resonances. 
The semi-analytical method, on the other hand, includes these resonances; notice, 
for instance, the small, weaker peaks at higher frequencies to the right of the 
fundamental resonance. These, however, carry little spectral weight and can only be seen due to the rather small 
damping parameter ($\Gamma=2.6$ meV). This justifies their omission when using the full-analytical technique, 
as their contribution is rather small.

\subsection{Ansatz for the current}
\label{appx:ansatz}

Here, we compare the ansatz for the current -- cf. Eq. (\ref{eq:Jx_ansatz}) -- against the same quantity computed as a Fourier 
series, as in the semi-analytic method [see Eq. (\ref{eq:sigmaCoeff})]. Notice that the overall correspondence is quite good, thus 
providing further evidence for the validity of the edge condition in our fully-analytic framework.
\begin{figure}[h]
  \centering
    \includegraphics[width=0.45\textwidth]{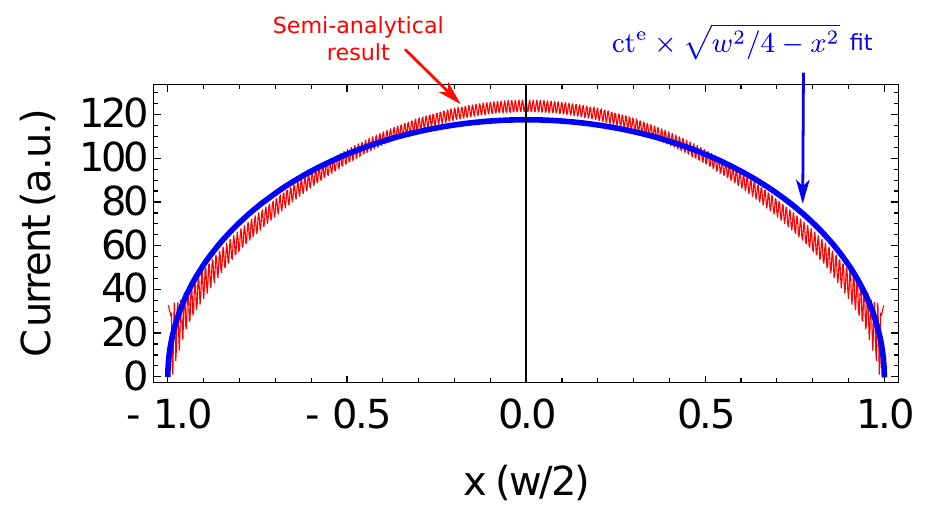}
      \caption[Ansatz for the current]{
      Current, $J_x(x)$, within the graphene stripes obtained using the semi-analytical method (Fourier expansion) 
      and corresponding fitting function of the type $\mathrm{ct}^{\mathrm{e}} \times \sqrt{w^2/4 - x^2}$ to illustrate 
      the approximation which is made when employing the edge condition [ansatz for the current; see Eq. (\ref{eq:Jx_ansatz})].}
    \label{SIfig:sqrt_fit}
\end{figure}
The oscillations emerging in the plot corresponding to the semi-analytic method are a natural consequence of the use of a Fourier 
expansion to describe the current in such geometry --- a feature that is known as Gibbs phenomenon \cite{Arfken}. 
Naturally, these are absent in the analytic ansatz for the current.

% :::::::::::::::::::::::::::::::::::::::::::::::::::::::::::::::::::::::::::::::::::::::::::::::::
% :::::::::::::::::::::::::::::::::::::::::::::::::::::::::::::::::::::::::::::::::::::::::::::::::

\section{Dynamical conductivity of graphene}
\label{appx:sigma_g}

In this work we model the dynamical conductivity of the graphene ribbons using Kubo's formula 
within the local approximation at room temperature ($T=300$ K). In such framework, the material's 2D conductivity reads \cite{Falkovsky,GoncalvesPeres}
\begin{align}
 \sigma_{\mathrm{Kubo}} (\omega) &= \sigma_{\mathrm{intra}} (\omega) + \sigma_{\mathrm{inter}} (\omega)\ , \label{def:sig_2cntr}\\
 \sigma_{\mathrm{intra}}(\omega) &= \frac{\sigma_0}{\pi} \frac{4}{\Gamma-i\hbar\omega}
 \left[ E_F + \frac{2}{\beta} \ln\left( 1 + e^{-\beta E_F} \right) \right]\ , \label{def:intra}\\
 \sigma_{\mathrm{inter}}(\omega) &= \frac{\sigma_0}{\pi} \left[ \pi G(\hbar\omega/2) \right. \nonumber\\
 &\quad + \left. i 4 \hbar \omega
 \int_{0}^{\infty} dE \frac{G(E) - G(\hbar\omega/2)}{(\hbar\omega)^2 - 4 E^2} \right]\ , \label{def:inter}
\end{align}
where $\beta= (k_B T)^{-1}$ (here $k_B$ is Boltzmann's constant), $\sigma_0 = e^2/(4\hbar)$, and where the quantity $G(x)$ is defined as
\begin{equation}
 G(x) = \frac{\sinh\left(x \beta\right)}{\cosh\left(E_F \beta\right) + \cosh\left( x \beta \right)}\ . \label{def:G}
\end{equation}

In the THz regime, for graphene under typical doping levels --- such that $E_F \gg k_B T$ and $2 E_F > \hbar\omega$ ---, the conductivity 
of graphene can be well approximated by the Drude-like expression
\begin{equation}
 \sigma_{\mathrm{D}} (\omega) \approx \frac{\sigma_0}{\pi} \frac{4 E_F}
{\Gamma-i\hbar\omega}\,,
\label{def:sig_limit}
\end{equation}
provided that the conditions $E_F \gg k_B T$ and $2 E_F > \hbar\omega$ are met.

% *********************************************************************************************
\section{Semi-analytical method in a nutshell}
\label{appx:semi-ana}

Similarly to the fully-analytical method 
described in the main text, the semi-analytical method also expresses the EM fields 
in the form of Bloch-sums.\cite{JPCM24,PolCrysPRB85,JOpt15,Primer,GoncalvesPeres} 
Namely, the fields in the medium $j$ may be written as 
(under normal incidence)
\begin{align}
 \mathcal{E}_x^{(j)}(x,z) &= E_x^{inc} e^{ik_z z} \delta_{j,1} + \sum_n E_{x,n}^{(j)} e^{inGx-\xi_{j,n} |z|}\ ,\label{FF:Exj}\\
 \mathcal{E}_z^{(j)}(x,z) &= E_z^{inc} e^{ik_z z} \delta_{j,1} + \sum_n E_{z,n}^{(j)} e^{inGx-\xi_{j,n} |z|}\ ,\label{FF:Ezj}\\
 \mathcal{B}_y^{(j)}(x,z) &= B_y^{inc} e^{ik_z z} \delta_{j,1} + \sum_n B_{y,n}^{(j)} e^{inGx-\xi_{j,n} |z|}\ ,\label{FF:Byj}
\end{align}
where $G=2\pi/L$, $k_z=\sqrt{\ep_1}k_0$ and $\xi_{j,n}^2 = (nG)^2 - \ep_j k_0^2$. Imposing the 
adequate boundary conditions, one obtains the following system of equations
\begin{align}
 Q_0 E_{x,0}^{(2)} + \frac{i}{\omega \epsilon_0}
 \sum_l \tilde{\sigma}_{-l} E_{x,l}^{(2)} &= i \frac{2 \epsilon_1}{k_z} E_x^{inc}\ ,\label{eq:LS1}\\
  Q_n E_{x,n}^{(2)} + \frac{i}{\omega \epsilon_0}
 \sum_l \tilde{\sigma}_{n-l} E_{x,l}^{(2)} &= 0\ .\label{eq:LS2}
\end{align}
for $n=0$ and $n \neq 0$, respectively, and where $Q_n = \epsilon_1 / \xi_{1,n} + \epsilon_2 / \xi_{2,n}$. In Eqs. (\ref{eq:LS1}) 
and (\ref{eq:LS2}), the quantities $\tilde{\sigma}_{m}$ are the components of the Fourier series that incorporates the system's 
periodicity, that is
\begin{align}
 \sigma(x) &= \sum_m \tilde{\sigma}_{m} e^{imGx}\ , \label{eq:sigma}\\
 \tilde{\sigma}_{m} &= \frac{1}{L} \int_{-L/2}^{L/2} \sigma(x) e^{-imGx} dx\ . \label{eq:sigmaCoeff}
\end{align}

The numerical solution of the (truncated) system of equations posed by Eqs. (\ref{eq:LS1}) and (\ref{eq:LS2}) for each 
frequency $\omega$ (entering as a parameter), renders the field amplitudes, $E_{x,l}^{(2)}$, in terms of $E_x^{inc}$. As before, only the 
mode with $n=0$ is propagating, and thus only this contributes (i.e. reaches the far-field) to the transmittance, reflectance and absorbance, which read

\begin{align}
 \mathcal{T}(\omega) &= \frac{\Re\{ \sqrt{\epsilon_2} \} }{\Re\{ \sqrt{\epsilon_1} \}}
 \left| \frac{E_{x,0}^{(2)}}{E_x^{inc}} \right|^2\ ,\label{eq:T}\\
 \mathcal{R}(\omega) &= \left| \frac{E_{x,0}^{(2)} - E_x^{inc}}{E_x^{inc}} \right|^2\ ,\label{eq:R}\\
  \mathcal{A}(\omega) &= 1 - \mathcal{T}(\omega) - \mathcal{R}(\omega)\ . \label{eq:A}
\end{align}
From these expressions, the corresponding spectra akin to the semi-analytical model may be readily obtained.

% -------------------------------------------------------------
% #                         References                        #
% -------------------------------------------------------------

%\bibliography{references}% Produces the bibliography via BibTeX.


\end{document}